\input harvmac 
\input tables
\input epsf
\def\figin{\epsfcheck\figin}\def\figins{\epsfcheck\figins}
\def\epsfcheck{\ifx\epsfbox\UnDeFiNeD
\message{(NO epsf.tex, FIGURES WILL BE IGNORED)}
\gdef\figin##1{\vskip2in}\gdef\figins##1{\hskip.5in}% blank space instead
\else\message{(FIGURES WILL BE INCLUDED)}%
\gdef\figin##1{##1}\gdef\figins##1{##1}\fi}
\def\DefWarn#1{}
\def\figinsert{\goodbreak\midinsert}
\def\ifig#1#2#3{\DefWarn#1\xdef#1{fig.~\the\figno}
\writedef{#1\leftbracket fig.\noexpand~\the\figno}%
\figinsert\figin{\centerline{#3}}\medskip\centerline{\vbox{\baselineskip12pt
\advance\hsize by -1truein\noindent\footnotefont{\bf Fig.~\the\figno:} #2}}
\bigskip\endinsert\global\advance\figno by1}

%Uses tables.tex, available from hepth via ``get tables.tex''
\font\zfont = cmss10 %scaled \magstep1
\def\ZZ{\hbox{\zfont Z\kern-.4emZ}}

\def\nc{N{_{c}}}

\def\nfl{{N_{f}}}\def\nfr{{N'_{f}}}
\def\ncl{{\nc}}\def\ncr{{N'_{c}}}
\def\ncld{{\tilde{N}_c}}\def\ncrd{{\tilde{N}'_c}}

\def\pf{{\rm Pf}\ }
%
% field defs

\def\Ql{Q}\def\Qr{Q'}

\def\Qlt{\tilde{\Ql}}\def\Qrt{\tilde{\Qr}}

%
% dual defs

\def\ql{q}\def\qr{\ql'}
\def\qlt{\tilde\ql}\def\qrt{\tilde\qr}

\def\np#1#2#3{Nucl. Phys. {\bf B#1} (#2) #3}
\def\pl#1#2#3{Phys. Lett. {\bf #1B} (#2) #3}

\lref\seiberg{N. Seiberg, {\it Electric-Magnetic Duality in Supersymmetric 
Non-Abelian Gauge Theories}, hep-th/9411149, \np {435}{1995}{129}.}

\lref\kutsch{D. Kutasov,
{\it A Comment on Duality in $N=1$ Supersymmetric Non-Abelian Gauge Theories},
\pl{351}{1995}{230}, hep-th/9503086; D. Kutasov and A. Schwimmer, 
{\it On Duality in Supersymmetric Yang-Mills Theory}, hep-th/9505004, 
\pl{354} {1995} 
315.}

\lref\kinsso{K. Intriligator and N. Seiberg, {\it Duality,
Monopoles,  Dyons, Confinement and Oblique Confinement in
Supersymmetric $SO(N_c)$ Gauge Theories},
hep-th/9503179, \np{444} {1995} {125}.}

\lref\kippsp{K. Intriligator and P. Pouliot, {\it Exact
Superpotentials, Quantum Vacua and Duality in Supersymmetric
$Sp(N_c)$ Gauge Theories}, hep-th/9505006, \np{353}
{1995} {471}.}

\lref\kenisosp{K. Intriligator, {\it New RG Fixed Points and
Duality in Supersymmetric $Sp(N_c)$ and $SO(N_c)$ Gauge
Theories}, hep-th/9505051, \np{448}{1995} 187.}

\lref\emop{R.G. Leigh and M.J. Strassler, {\it Exactly Marginal
Operators and Duality in N=1 Supersymmetric Gauge Theories}, 
hep-th/9503121, \np{447}{1995}{95}.}

\lref\rlmssosp{R.G. Leigh and M.J. Strassler, {\it Duality of
$Sp(2N_c)$ and $SO(N_c)$ Supersymmetric Gauge Theories with
Adjoint Matter}, hep-th/9505088, \pl{356}{1995} 492.}

\lref\ilslist{K. Intriligator, R.G. Leigh and M.J. 
	Strassler, {\it New Examples 
of Duality in Chiral and Non-Chiral Supersymmetric Gauge Theories},
hep-th/9506148, \np{456}{1995}567.}

\lref\spinduals{P. Pouliot, {\it Chiral Duals of Non-Chiral SUSY Gauge 
Theories}, hep-th/9507018, \pl{359} {1995} 108;
		P. Pouliot and M.J. Strassler, {\it A Chiral $SU(N)$ Gauge 
Theory and its Non-Chiral $Spin(8)$ Dual}, hep-th/9510228,
		\pl{370} {1996} 76;
                P. Pouliot and M.J. Strassler, {\it Duality and Dynamical 
Supersymmetry Breaking in $Spin(10)$ with a Spinor}, hep-th/9602031,
		\pl{375} {1996} 175; T. Kawano, {\it Duality of N=1 
Supersymmetric SO(10) Gauge Theory with Matter in the Spinorial 
Representation},  hep-th/9602035, Prog. Theor. Phys. {\bf 95} (1996) 963.}

\lref\brodie{J. Brodie, {\it Duality in Supersymmetric $SU(N_c)$ Gauge 
Theory with Two Adjoint Chiral Superfields}, PU--1626, hep-th/9605232,
\np{478}{1996}{123}.}

\lref\singularity{V.I. Arnold, {\it Singularity Theory}, 
London Mathematical Lecture Notes Series: 53, 
Cambridge University Press (1981).}

\lref\berkooz{M. Berkooz, {\it The Dual of Supersymmetric SU(2k)
with an Antisymmetric Tensor and Composite Dualities}, hep-th/9505067,
\np{452}{1995} 513; P. Pouliot, {\it Duality in SUSY $SU(N)$ with an
Antisymmetric Tensor}, hep-th/9510148,\pl{367}{1996} 151.}

\lref\otherdeconf{Markus A. Luty, Martin Schmaltz, and John Terning, 
{\it A Sequence of Duals for $Sp(2N)$ Supersymmetric Gauge Theories with 
Adjoint Matter}, UMD-PP-96-71, hep-th/9603034, to be published in
{\sl Phys. Rev. D}.}

\lref\kss{D. Kutasov, A. Schwimmer, and N. Seiberg, {\it Chiral Rings, 
Singularity Theory and Electric-Magnetic Duality}, hep-th/9510222, 
\np{459}{1996} 455.}

\lref\csakietal{C. Csaki, M. Schmaltz, and W. Skiba, {\it Exact Results and 
Duality for $Sp(2N)$ SUSY Gauge Theories with an Antisymmetric
Tensor}, MIT-CTP-2552, hep-th/9607210.}

\lref\negdim{R. Penrose, 
{\it Combinatorial Mathmatics and its Applications} 
(1971) p221. ed. D. Welsh (New York Academic Press); 
R. C. King, Can. J. Math. {\bf 23} (1971) 176; P. Cvitanovic and A.D. Kennedy,
  Phys. Scr. {\bf 26} (1982) 5; G.V. Dunne, J. Phys. A: Math. Gen. {\bf 22} 
(1989) 1719; G. Parisi and N. Sourlas, Phys. Rev. Lett.
{\bf 43} (1979) 744;N. Maru and S. Kitakado, {\it Negative dimensional group 
extrapolation and dualities in N = 1 supersymmetric gauge theories}, 
DPNU-96-52, hep-th/9609230.}

\lref\M{E. Witten, {\it Bound States Of Strings And $p$-Branes},
hep-th/9510135, \np{460}{1996}{335};
T. Banks, W. Fischler, S. H. Shenker, L. Susskind, {\it 
M Theory As A Matrix Model: A Conjecture}, RU-96-95, SU-ITP-96-12, 
UTTG-13-96, hep-th/9610043.}

\lref\distler{J. Distler, A. Karch, {\it N=1 Dualities for Exceptional 
Gauge Groups and Quantum Global Symmetries}, UTTG-20-96, hep-th/9611088;
P. Ramond, {\it Superalgebras in N=1 Gauge Theories}, UFIFT-HEP-96-19, 
hep-th/9608077.}

%\lref{}

\Title{\vbox{\rightline{hep-th/9611197}\rightline{PU--1658 }
\rightline{IASSNS--HEP--96/110}}}
{\vbox{\centerline{Patterns of Duality in N=1 SUSY Gauge Theories}}}
\centerline{{or:} 
%}
%\centerline{
%``
{\it Seating Preferences of Theatre-Going Non-Abelian Dualities}
%''
} 
\bigskip\medskip
\centerline{ John H. Brodie}
\smallskip
{\it
\centerline{Department of Physics}
\centerline{Princeton University}
\centerline{Princeton, NJ 08540, USA}}
\centerline{\tt jhbrodie@princeton.edu}
\bigskip\medskip
\centerline{Matthew J. Strassler}
\smallskip{\it
\centerline{School of Natural Sciences}
\centerline{Institute for Advanced Studies}
\centerline{Princeton, NJ 08540, USA}}
\centerline{\tt strasslr@ias.edu}

%\vskip .2in

\vglue .3cm

\noindent
We study the patterns in the duality of a wide class of N=1
supersymmetric gauge theories in four dimensions.  We present many new
generalizations of the classic duality models of Kutasov and
Schwimmer, which have themselves been generalized numerous times in
works of Intriligator, Leigh and the present authors.  All of these
models contain one or two fields in a two-index tensor representation,
along with fields in the defining representation.  The superpotential
for the two-index tensor(s) resembles $A_k$ or $D_k$ singularity
forms, generalized from numbers to matrices.  Looking at the ensemble
of these models, classifying them by superpotential, gauge group, and
``level'' --- for terminology we appeal to the architecture of a
typical European-style theatre --- we identify emerging patterns and
note numerous interesting puzzles.
\Date{11/96}

\newsec{Introduction}

At present the known duality transformations in 
four-dimensional N=1 supersymmetric gauge
theories are few, and generally  fall into a few special classes.  The
first examples originated in the work of Seiberg \seiberg\ and were
elaborated in Refs.~\kinsso\ and \kippsp.  These dualities exchange two
theories with a single gauge group $SU(N)$, $SO(N)$ or $Sp(N)$, and
matter in the defining representation of the gauge group (fundamental
plus anti-fundamental for $SU$, vector for $SO$, fundamental for $Sp$.)
Certain gauge singlets are also required.

  In the models of Ref.~\kutsch, a field $X$ in the adjoint
representation is added to the $SU$ duality of Ref.~\seiberg.  This
field is given a superpotential $\Tr X^{k+1}$, corresponding to an $A_k$
singularity under the usual ADE classification \singularity.  The dual
theory also has an adjoint field $\bar X$ with superpotential
$\Tr \bar X^{k+1}$.  The duality reduces to that of \seiberg\ in the
case $k=1$.  This $A_k$ singularity structure can be generalized to a
wide variety of other gauge groups and matter content, as in
Refs.~\refs{\kenisosp,\rlmssosp, \ilslist}.

Additional and unrelated dualities, involving spinors of $Spin$ groups
and their subgroups, were found in \spinduals, and were shown to flow
to the $SO$ duality of Refs.~\refs{\seiberg,\kinsso}.  Other claims of
duality transformations have been made in
\refs{\berkooz,\otherdeconf,\csakietal,\distler}; many of these follow from the
ones listed above.

Recently, one of us presented \brodie\ a new generalization of Ref.~\kutsch.
Instead of one adjoint field $X$ with an $A_k$ superpotential, two
adjoint fields $X$ and $Y$ were introduced and were given a $D_{k+2}$
superpotential of the form 
\eqn\Dkpot{
W=\Tr X^{k+1}+\Tr XY^2.} 
It was shown that for $k$ odd this model has the
special properties which have appeared in all $A_k$ dualities (namely,
its chiral ring is truncated) and is dual to a model of similar type.

Furthermore, Ref.~\brodie\ also presented a duality in which the gauge 
group is $SU(N_c)\times
SU(N_c')$, with a field $F$ in the ${\bf (N_c,N'_c)}$ representation,
another $\tilde F$ in the conjugate representation, fields $X_1$ and
$X_2$ in the adjoint representations of the two group factors, and
for each group factor a number of flavors in its fundamental and
anti-fundamental representations.  The superpotential is
\eqn\AkAkpot{
W=\Tr X_1^{k+1}+ \Tr X_2^{k+1} +\Tr  X_1{\tilde F} F -\Tr  X_2 {\tilde F} F}
and again the theory is dual to a model of similar type.

In this paper, we present a number of generalizations of the two
classes of dual pairs presented in Ref.~\brodie. We quickly review the
$A_k$ models and outline the pattern of the $D_{k+2}$ models. 
After touching on the key points of the $D_{k+2}$ model 
of Ref.~\brodie, we
examine new $D_{k+2}$ models with $SO(N_c)$ gauge symmetry with two
symmetric tensors and $Sp(N_c)$ gauge symmetry with two anti-symmetric
tensors. We review the $SU(N_c)\times SU(N_c')$ duality of Ref.~\brodie, and
derive it from the $D_{k+2}$
model using an improved and simplified method.  
We then generalize it 
to $SO\times SO$ and $Sp\times Sp$
groups.  Next, we examine a large number of generalizations of the
$D_{k+2}$ class, involving $SU$, $SO$ and $Sp$ gauge groups.  We
consider various representations for the field $Y$, but with $X$
always in the adjoint (symmetric) [anti-symmetric] representation for
$SU$ ($SO$) [$Sp$] groups.  In each of these cases, the $D_3$ models are the 
same as the $A_3$ models of 
Refs.~\refs{\kutsch,\kenisosp,\rlmssosp,\ilslist}.  
Finally, we discuss the
overall structure of the known examples and list puzzles which remain
unsolved at the present time.

\newsec{Review of $A_k$ Models}

\ifig\figAk{
The simplest $A_k$ models of 
Refs.~\refs{\kutsch,\kenisosp,\rlmssosp, \ilslist}, organized as
suggested by Sec.~2 of \ilslist.  The notation in the figure is as
follows: $X$ represents a field in the adjoint representation of
$SU(N)$, $S$ and $A$ represent symmetric and anti-symmetric tensor
representations, and a tilde represents a conjugate representation of
$SU(N)$.  For $SU$ $[SO]$ $(Sp)$ groups, $N=N_c$ $[N_c]$ $(2N_c)$ and 
$F=N_f$ $[N_f]$ $(2N_f)$. The superpotential for each model
is given.  Each model is dual to a model of similar type, with color
group $SU(\tilde N)$, $SO(\tilde N)$ or $Sp({\tilde N\over 2})$.  The arrows
indicate that there are flat directions along which the upper models
flow to the lower ones.}
{\epsfxsize4.0in\epsfbox{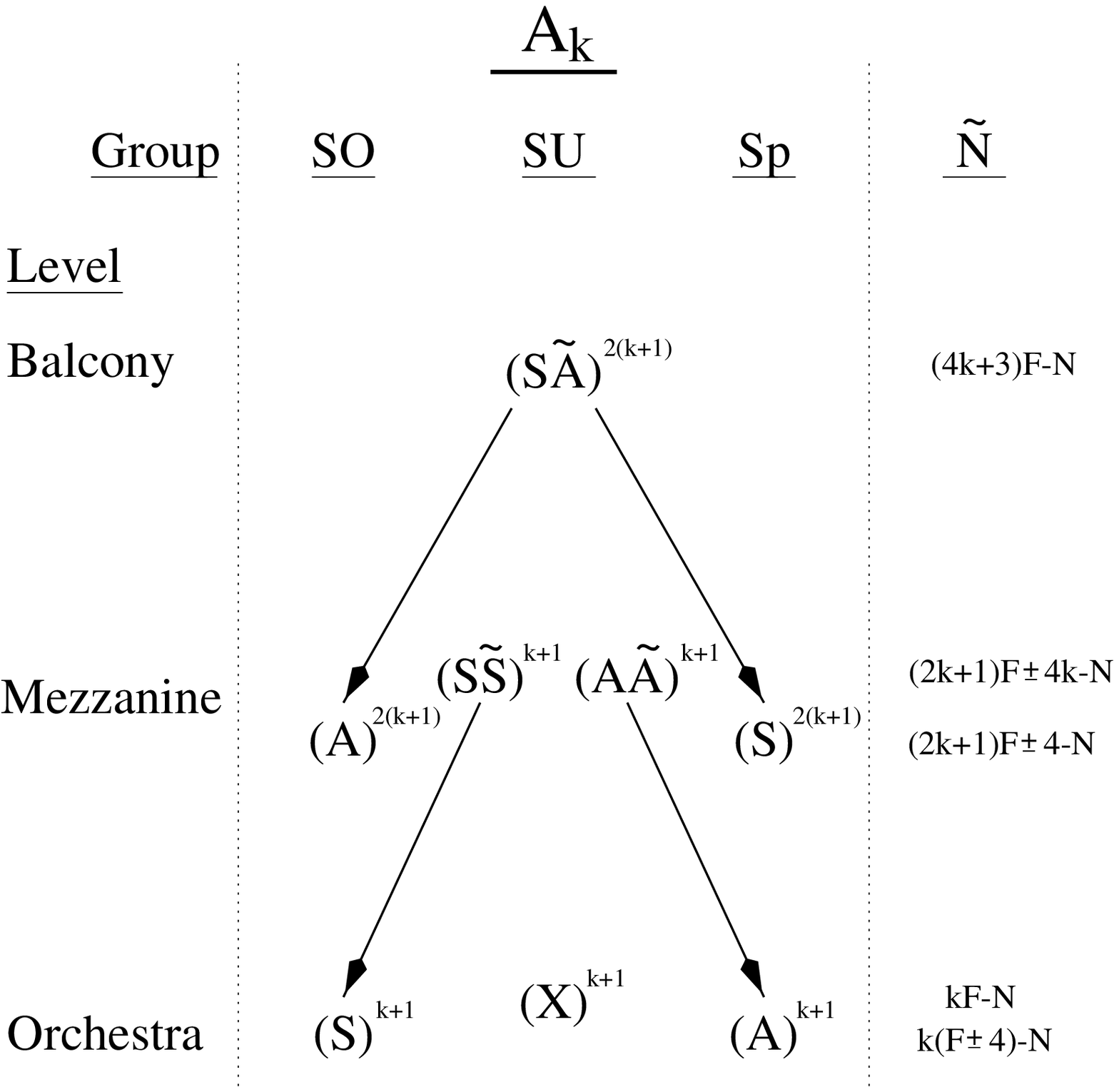}}

\ifig\figAkloge{
The models of the previous figure along with the
loge models of Ref.~\ilslist.  The loge models consist of a product
of two or three gauge groups, which is displayed on the figure.  
Each is dual to a model of similar type, with the same groups.
The process of confinement of one group factor causes the
loge models to flow along the arrows in the diagram to models
with one fewer group factor.  More generally, duality of a single group
factor relates the duality of each loge model to a duality
of a model with one fewer group factor, along the same arrows.}
{\epsfxsize4.0in\epsfbox{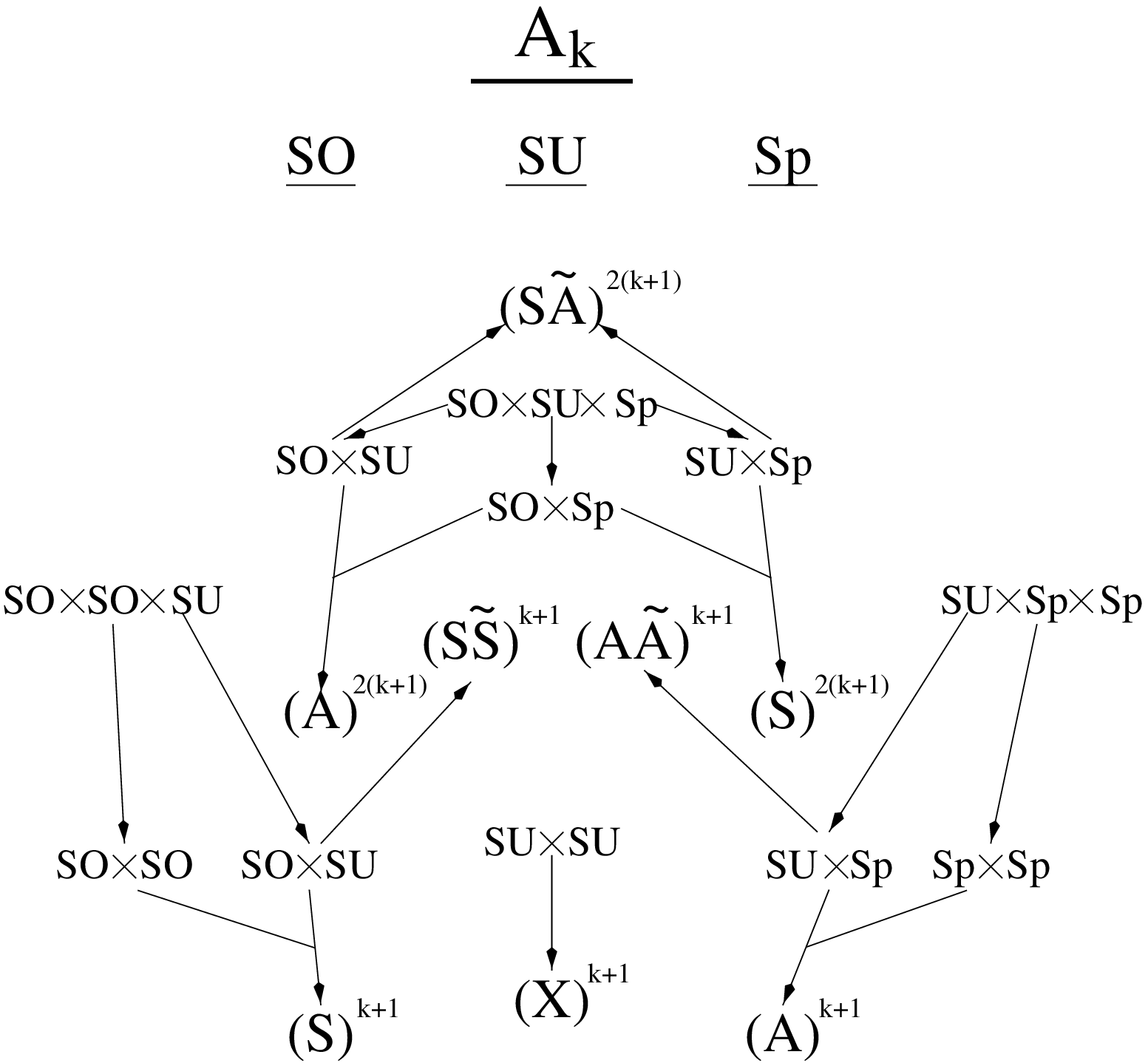}}

We begin by reviewing the results of
Refs.~\refs{\kutsch,\kenisosp,\rlmssosp,\ilslist}. In that series of
papers, many classes of $A_k$-type models were exhibited; a summary of
those nineteen models is given in Sec.~2.1--2.19 of \ilslist.  They
can be organized into the multi-level structure of
\figAk\ and \figAkloge.  For ease of reference we have chosen to refer 
to them using the terminology for the levels of a standard European-style
theatre.

The orchestra models \refs{\kutsch,\kenisosp} have a superpotential of
the form $\Tr X^{k+1}$, where $X$ is in the adjoint (symmetric)
[anti-symmetric] representation for $SU$ ($SO$) [$Sp$] groups.  In
addition each model has a number of fields $Q$ in the defining
representation of the gauge group, transforming under a flavor
symmetry.  Each model is dual to a theory of similar type, containing
in addition certain fields $M$ which are singlets of the gauge group
but non-singlets of the flavor symmetry.  The dual superpotential for
the dual field $\bar X$ is again $\Tr \bar X^{k+1}$ and, in addition,
contains terms coupling the mesons $M$ to the dual fields $q$ and
$\bar X$.

The mezzanine models \refs{\rlmssosp,\ilslist} 
differ from the orchestra models in that they have a superpotential
$\Tr X^{2(k+1)}$ for $SO$ ($Sp$) with $X$ in the anti-symmetric
(symmetric) representation, or $\Tr (X\tilde X)^{k+1}$ for $SU$ with
$X$ in the symmetric or anti-symmetric representation and $\tilde X$
in its conjugate.

The balcony model, which is chiral, has gauge group $SU$ and a superpotential 
$\Tr (X\tilde X)^{2(k+1)}$, 
where $X$ is in the symmetric and $\tilde X$ is in the 
conjugate anti-symmetric representation.

These models and their interrelations are shown in 
\figAk.
Notice the many patterns in the dual color groups, listed at the right
of the figure.  The duality transformations are
invariant under simultaneous exchange of $SO$ and $Sp$ groups and
symmetric and anti-symmetric tensors, along with $N\leftrightarrow -N$,
$\tilde N\leftrightarrow -\tilde N$,
and $F\leftrightarrow -F$.  The group theory behind this
$\ZZ_2$ symmetry has been discussed in the literature \negdim .

Some of these models have flat directions along which
they flow to others in the infrared, as indicated by the arrows in 
\figAk. An important check on the conjectured duality transformations is that
they commute with these flows  \ilslist\ in a highly non-trivial way.

Related to each of these models are the loge models \ilslist\ shown in 
\figAkloge\ in which
the fields $X$ or $\tilde X$ are deconfined.  An adjoint
representation of $SU(N)$ can always be deconfined
\refs{\seiberg,\berkooz,\ilslist,\otherdeconf} using an $SU(N)\times
SU(N-1)$ group with a field $F$ in the $({\bf N,N-1})$ and a field
$\tilde F$ in its conjugate.  Symmetric [anti-symmetric]
representations of a group $G$ can always be deconfined using
$SO(N)\times SO(N+4)$ [$Sp(N)\times Sp(N-2)$] groups with a single
field in the $({\bf N,N+4)}$ [$({\bf N,N-2)}$].  Under this
deconfinement the degree of the superpotential is increased, since
each field $X$ is rewritten as $FF$ or $F\tilde F$.  What is
remarkable is that, in contrast to various other examples of
deconfinement which appear in the literature
\otherdeconf, the dualities of Refs.~\refs{\kutsch,\kenisosp,\rlmssosp,
\ilslist,\spinduals} are preserved when deconfinement is applied to 
both sides; a complete closed diagram of theories, in which two dual
loge models flow to two dual orchestra, mezzanine, or balcony models,
is generated.

\vglue0.25in

\ifig\figclose{
A closure diagram for a dual pair of theories.
This is a crucial consistency test of any claim of duality.  
The two gauge groups at the top of the diagram are supposed to
describe the same theory in the infrared,  
Ref.~\ilslist.  This should be true for any choice of
gauge couplings.  Take the coupling of $SO(N)$ to be much larger than
that of $Sp(M)$; the duality requires the same for $SO(\tilde N)$.
A strongly coupled $SO(N)$ theory is better described by its dual;
here the appropriate dual is that of Ref.~\seiberg.  The low-energy
$SO(\hat N)$ theories have the same color group and are
weakly coupled spectators to the $Sp$ groups, which fortunately
are related by a duality of Ref.~\rlmssosp.}
{\hglue0.5in{\epsfxsize3.0in\epsfbox{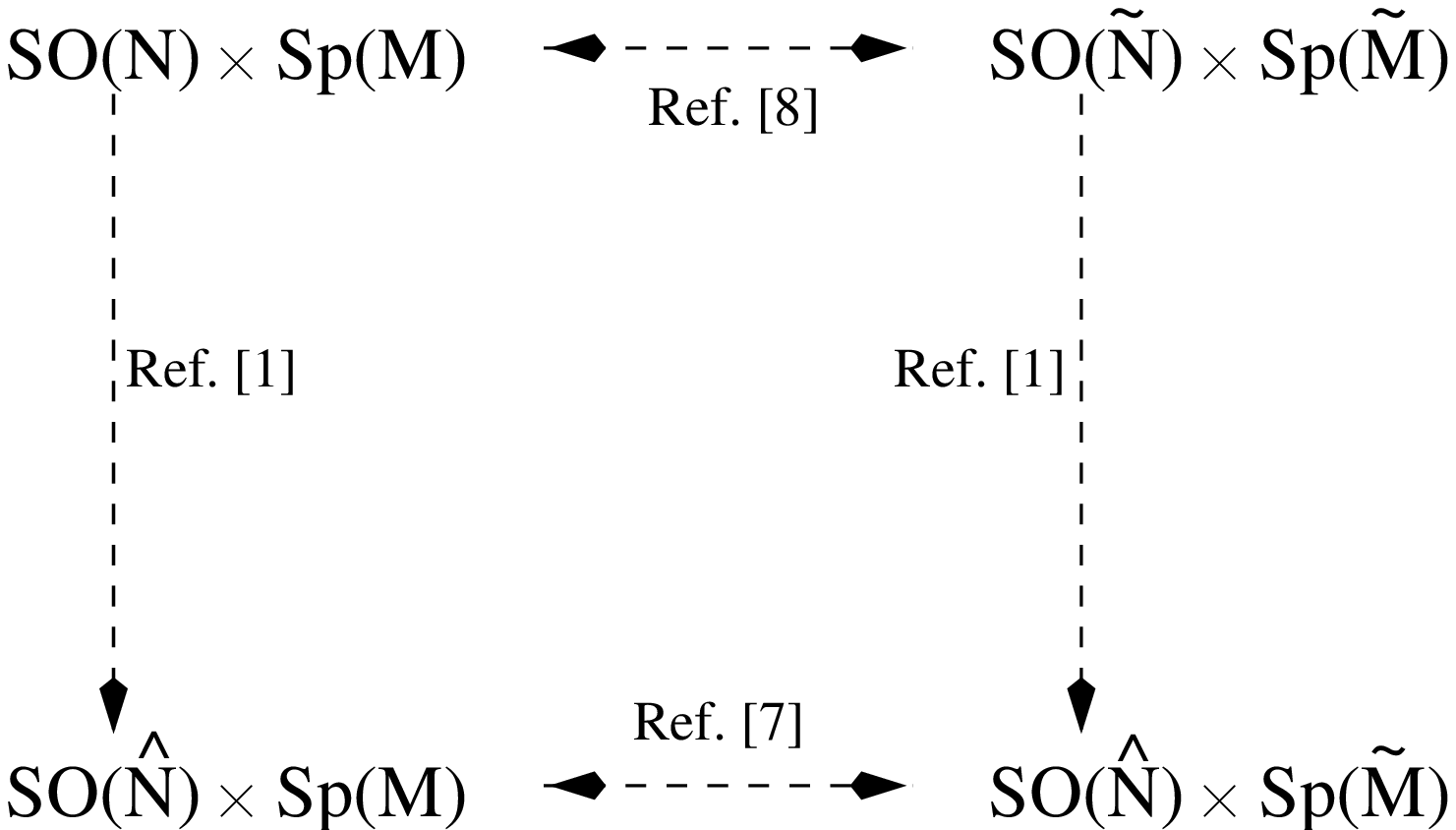}}}

An example of this process is illustrated in \figclose . Consider the 
theory built on
$SO(N)\times Sp(M)$, with a field $F$ in the ${\bf(N,2M)}$, $N_f$
fields $Q$ in the vector representation of $SO(N)$ and $M_f$ fields in
the fundamental representation of $Sp(M)$.  The superpotential is
$\Tr F^{4(k+1)}$. As described in Sec.~2.12 of \ilslist, the magnetic
theory is similar, with gauge group $SO(\tilde N)\times Sp(\tilde M)$ where
\eqn\SOSpdual{\eqalign{\tilde N = & 2(k+1)(M_f+N_f)-N_f-2M\ ; \cr 
2\tilde M = & 2(k+1)(M_f+N_f)-M_f-N\ .\cr} } If the $SO(N)$ group
becomes strongly coupled at an energy scale where the $Sp(M)$ group is
still weakly coupled, then a better description of the theory below
that scale is given by using the dual of the $SO(N)$ group (which has
$N_f+2M$ vector representations) with respect to the duality of
Seiberg \refs{\seiberg,\kinsso}.  The theory then has gauge group
$SO(\hat N)\times Sp(M)$, where $\hat N=N_f+2M-N+4$, and the $Sp$
factor now has a symmetric field $S=F^2$ and a field $f$ in the
${\bf(\hat N,2M)}$ with superpotential $\Tr S^{2(k+1)} + Sff$, along
with $M_f$ fundamentals, which makes it an $A_k$ mezzanine theory.
Meanwhile, in the dual $SO(\tilde N)\times Sp(\tilde M)$ theory, the
$SO(\tilde N)$ theory becomes strongly coupled first and is also best
represented by its dual under the duality of Seiberg.  The low-energy
theory is then $SO(\hat{\tilde N})\times Sp(\tilde M)$, where
$\hat{\tilde N}=M_f+2\tilde M-\tilde N+4=\hat N$, from \SOSpdual.  The
$Sp$ factor is now also a mezzanine theory.  One may check, counting
carefully the number of fundamentals in the $Sp$ groups and accounting
for the $Sff$ term in the superpotential, that the $Sp(M)$ and
$Sp(\tilde M)$ factors are indeed dual under the mezzanine duality of
\rlmssosp.

If instead the $Sp$ factors had become strongly coupled
before the $SO$ factors, the low-energy duality would have
involved the $SO$ mezzanine models.

Since each of the $A_k$ dualities relates two theories of similar type,
there are choices of $N_f$ and $N_c$ such that the theories are
self-dual; that is, the electric and magnetic theories have the same
gauge group and charged matter, though they will differ through the
presence or absence of gauge singlet fields.  
The self-dual theories of Seiberg \seiberg\ have exactly marginal
operators which take the form of meson mass terms and which
are deeply connected with the duality transformation \emop.  The
presence of marginal operators in the self-dual models of the $A_k$
series was noted in a few examples \refs{\kutsch,\rlmssosp} and shown to be
ubiquitous in Ref.~\ilslist.  The precise connection with the duality
transformation is as yet unknown.

\newsec{A Program Guide for the $D_{k+2}$ Models}

\ifig\figDk{
The ensemble of $D_{k+2}$ models described in this
paper, along with the product models to which they are
related.  Notation is similar to \figAk, with $X,Y$ adjoint fields
of $SU(N)$, $S,T$ symmetric tensors and $A,B$ anti-symmetric tensors,
and a tilde representing a conjugate representation of $SU(N)$. For 
$SU$ $[SO]$ $(Sp)$ groups, $N=N_c$ $[N_c]$ $(2N_c)$ and $F=N_f$ 
$[N_f]$ $(2N_f)$. The superpotential for each model is given,
with the first term listed at the top and the second at the position
of the model in the diagram; thus the superpotential for the Mezzanine
$SO(N)$ model is $S^{k+1}+SA^2$, {\it etc}.  Each model is dual to a
model of similar type, with color group $SU(\tilde N)$, $SO(\tilde N)$
or $Sp({\tilde N\over 2})$.  The dashed arrows indicate that under certain 
perturbations the $D_{k+2}$ models flow to stage models, except for the 
balcony model which flows to copies
of itself with lower $k$.}
{\hglue0.5in{\epsfxsize3.6in\epsfbox{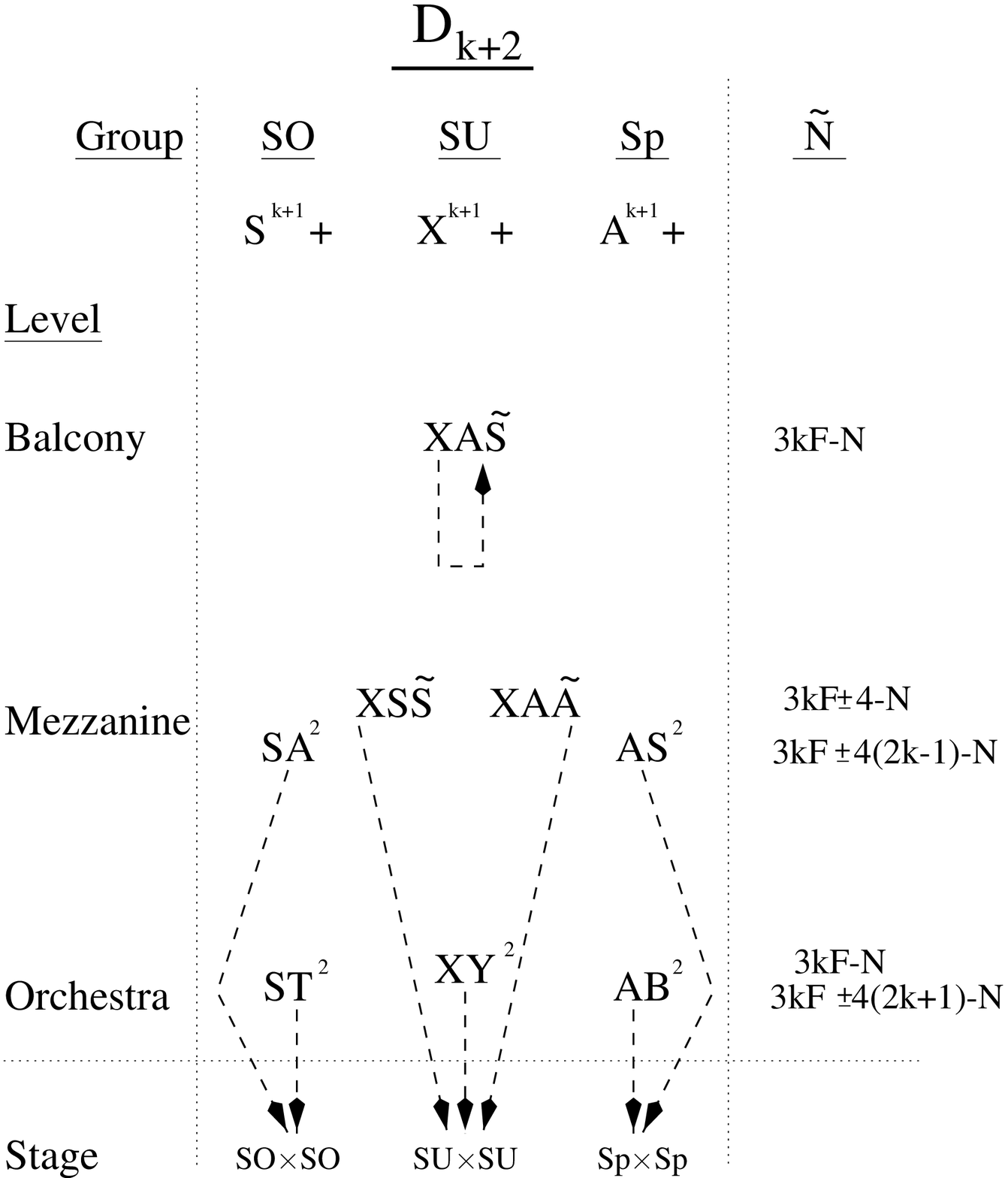}}}

We present here a preview of the models we found that have a 
superpotential of the form $D_{k+2}$. 
We have organized them, like the $A_k$ models, into a multi-level structure 
(see \figDk). 

For all the models, we have a superpotential of the form $\Tr X^{k+1}
+ \Tr XY^2$ where $X$ is in the traceless adjoint (symmetric)
[anti-symmetric] representation for $SU$ ($SO$) [$Sp$] groups.  In the
orchestra models, the field $Y$ is also in the traceless adjoint
(symmetric) [anti-symmetric] representation for $SU$ ($SO$) [$Sp$]
groups.  In addition each model has a number of fields $Q$ in the
defining representation of the gauge group, transforming under a
flavor symmetry.  Each model is dual to a theory of similar type,
containing in addition certain fields $M$ which are singlets of the
gauge group but non-singlets of the flavor symmetry.  The dual
superpotential for the dual fields $\bar X, \bar Y$ is again $\Tr \bar
X^{k+1} + \Tr \bar X \bar Y^2$ and, in addition, contains terms
coupling the mesons $M$ to the dual fields $q$, $\bar X$ and $\bar Y$.

The mezzanine models differ from the orchestra models in that $Y$ is
in the anti-symmetric (symmetric) representation for $SO$ ($Sp$),
while for $SU$ the field $Y$ is in the anti-symmetric or symmetric
representation and $\tilde Y$ is in its conjugate.  In the latter case
the superpotential is $W = \Tr X^{k+1} + \Tr XY\tilde Y$.

The balcony model, which is chiral, has gauge group $SU$ with fields
$Y$ in the anti-symmetric and $\tilde Y$ in the conjugate symmetric
representation; again the superpotential is $W = \Tr X^{k+1} + \Tr
XY\tilde Y$.

In the $A_k$ models, we have examples of dualities involving product
gauge groups, which we call the loge models.  The dynamics of
these models can cause them to flow in the infrared to the $A_k$
models of 
\figAk, as shown in 
\figAkloge.  We have not found
analogues of the loge models for $D_{k+2}$.  Instead, we have found other
models with product gauge groups, generalizing that of
Ref.~\brodie, with superpotentials in the form of Eq.~\AkAkpot .
We will refer to these as stage models.
The $D_{k+2}$ orchestra and  mezzanine models flow to the stage models under
perturbation, as shown by the arrows in 
\figDk, preserving the duality.  

As in the $A_k$ models, all of these theories are dual to models of
similar type.  As before there is a symmetry under which $N$, $\tilde
N$ and $F$ change sign, $SO$ and $Sp$ groups are interchanged, and
symmetric and anti-symmetric tensors representations are exchanged.
Also, each of the models has choices of $N_f$ and $N_c$ for which it
is self-dual and for which there are exactly marginal operators in the
form of meson mass terms.

\newsec{The $D_{k+2}$ Orchestra Models}

\subsec{Orchestra: $SU(N_c)$ with Two Adjoint Tensors}

In Ref.~\brodie\ a $D_{k+2}$ generalization of the $A_k$ models of
\kutsch\ was found for odd $k$.  The theory, with gauge group
$SU(N_c)$, has the following matter content:
\thicksize=1pt
\vskip12pt
\begintable
\tstrut  |$SU(\ncl)$  |$SU(\nfl)_L$|$SU(\nfl)_R$|$U(1)_B$|$U(1)_R$ \crthick
$\Ql$| ${\bf\ncl}$ |${\bf\nfl}$ | 
 ${\bf 1}$ | 1 | $1-{\ncl\over\nfl (k+1)}$   \cr
$\Qlt$| ${\bf\overline\ncl}$ |${\bf1}$ |
 ${\bf \overline\nfl}$ | $-1$ | $1-{\ncl\over\nfl (k+1)}$   \cr
$X$ | ${\bf\ncl^2 -1}$ | ${\bf 1}$ | ${\bf 1}$ | 0 |  ${2\over k+1}$    \cr
$Y$ | ${\bf\ncl^2 -1}$ | ${\bf 1}$ | ${\bf 1}$ | 0 |  ${k\over k+1}$
\endtable
\noindent
The superpotential is 
\eqn\AAb{W = {\Tr X^{k+1}\over k+1} + \Tr XY^2 + \lambda_1 \Tr X 
+ \lambda_2 \Tr Y}
where $\lambda_1$ and $\lambda_2$ are Lagrange multipliers which
enforce the tracelessness condition on $X$ and $Y$.  The conditions 
for a supersymmetric vacuum that follow from this superpotential are
\eqn\Aeom{\eqalign{X^k + Y^2 + \lambda_1 = 0\cr 
XY+YX + \lambda_2 = 0. \cr}} These equations truncate the chiral
ring for odd values of $k$. To illustrate this we can ignore
$\lambda_1$ and $\lambda_2$; statements below will then be correct modulo
lower order terms already included in the chiral ring.  We can
multiply $X^k$ by $Y$ from the right or left, and use the first
equation in \Aeom\ to show that
\eqn\Aring{YX^k + X^kY = -2Y^3.}
Now, we can use the second equation in \Aeom\ to anticommute the $Y$
fields through the $X$ fields.
\eqn\Bring{((-1)^k+ 1)X^kY = -2Y^3}
Thus for odd $k$, $Y^3=0$. The chiral ring is now said to be truncated
since it depends only on $k$ and not on the size of the gauge group.  
The gauge invariant mesons are
\eqn\Amesons{\tilde Q_s X^jY^l Q^r}
where $r$ and $s$ are flavor indices and 
where the truncation requires $j = 0,1, \cdots ,k-1$ and $l = 0,1,2$.
These mesons are all in the $(\bf {N_f,\overline N_f})$ representation
of the flavor group.  

This theory has a dual description in terms of an $SU(3kN_f - N_c)$ gauge
theory with the following charged matter fields
\thicksize=1pt
\vskip12pt
\begintable
\tstrut  |$SU(\ncld)$  |$SU(\nfl)_L$|$SU(\nfl)_R$|$U(1)_B$|$U(1)_R$ \crthick
$\ql$| ${\bf\ncld}$ |${\bf\overline\nfl}$ | 
 ${\bf 1}$ | ${N_c\over \tilde N_c}$ | $1-{\ncld\over\nfl (k+1)}$   \cr
$\qlt$| ${\bf\overline\ncld}$ |${\bf1}$ |
 ${\bf \nfl}$ | $-{N_c\over \tilde N_c}$ | $1-{\ncld\over\nfl (k+1)}$   \cr
$\bar X$ | ${\bf\ncld^2 -1}$ | ${\bf 1}$ | ${\bf 1}$ | 0 |  ${2\over k+1}$ \cr
$\bar Y$ | ${\bf\ncld^2 -1}$ | ${\bf 1}$ | ${\bf 1}$ | 0 |  ${k\over k+1}$
\endtable
\noindent
Throughout this paper, $\tilde N_c$ will be used to denote the dual
gauge group.  The dual description also has gauge singlets, $(M_{jl})^r_s$,
which are in a one-to-one mapping with the mesons in equation \Amesons.
The dual superpotential has the form
\eqn\Af{W={\Tr \bar X^{k+1}\over k+1} + \Tr \bar X \bar Y^2
+ \sum_{j=0}^{k-1} 
\sum_{l=0}^{2} M_{jl}\tilde q \bar X^{k-j-1} \bar Y^{2-l} q
+\lambda_1 \Tr \bar X 
+ \lambda_2 \Tr \bar Y \ .}
where flavor indices are contracted in the obvious way.

If we consider the case $k=1$ in \AAb, we see that the fields $X$ and
$\bar X$ are massive and can be integrated out in the IR. The
theories develop a quartic superpotential for $Y$, $\bar Y$, and the
low-energy theories are (marginal deformations of) models known to
be dual through the work of Kutasov and Schwimmer \kutsch.  This
is a manifestation of the equivalence between $D_3$ and $A_3$ 
singularities (though in this matrix generalization of singularity
theory the equivalence is not exact.)

The duality is consistent along certain flat
directions of this theory. Consider the superpotential \AAb\ where $k$
is taken to be odd.  The minima of the superpotential \Aeom\ allow for
an $SU(2n+km)$ gauge theory to have a flat direction along which $X$
and $Y$ get vacuum expectation values
\eqn\vevxy{\eqalign{
\vev{X}=&a\left(\matrix{0_n&0&0&0&0&0&.&0\cr
                  0&0_n&0&0&0&0&.&0\cr
                  0&0&1_m&0&0&0&.&0\cr
                  0&0&0&\omega_m &0&0&.&0\cr
                  0&0&0&0&\omega_m^2&0&.&0\cr
                  0&0&0&0&0&\omega_m^3&.&0\cr
		  .&.&.&.&.&.&.&.\cr
		  0&0&0&0&0&0&.&\omega_m^{k-1}\cr}\right)\cr\cr
\vev{Y}=&b\left(\matrix{1_n&0&0&0&0&0&.&0\cr
                  0&-1_n&0&0&0&0&.&0\cr
                  0&0&0_m&0&0&0&.&0\cr
                  0&0&0&0_m&0&0&.&0\cr
                  0&0&0&0&0_m&0&.&0\cr
                  0&0&0&0&0&0_m&.&0\cr
                  .&.&.&.&.&.&.&.\cr
                  0&0&0&0&0&0&.&0_m\cr}\right)}}
where $\omega=\exp{2\pi i\over k}$ and $a^k=b^2=\lambda_1$.  (A
subscript $n$ on an entry indicates that it is proportional to the
$n\times n$ dimensional unit matrix.)  The expectation value for $Y$
breaks the theory into $SU(n)\times SU(n)\times SU(km)\times U(1)^2$;
that of the $X$ field breaks it down to $SU(n)\times SU(n)\times
SU(m)^k\times U(1)^{k+1}$.  The fields $X$ and $Y$ each decompose into
$k+2$ adjoints and fields in the $({\bf n,n})$, $({\bf n,m})$, $({\bf
m,n})$, and $({\bf m,m})$ representations.  The field from $X$ in the
$({\bf n,n})$ representation, which we call $F$, and its conjugate
$\tilde F$, do not receive a mass, owing to the tracelessness
condition on $X$.  The leading term in the low-energy superpotential is
\eqn\intW{W_L = {\Tr (F\tilde F)^{k+1\over 2}\over k+1} \ .}
All the other matter fields from $X$ and $Y$ are massive along this
flat direction.  The same expectation values in the dual theory
give a similar result; the $SU(3kN_f - N_c)$ gauge group is broken to
$SU(kN_f-n)\times SU(kN_f-n)\times SU(N_f-m)^k\times U(1)^{k+1}$.  
The $SU(n)\times SU(n)$ factors of the low-energy theories are related
under an $A_k$ loge duality discussed in Sec.~2.11 of Ref.~\ilslist;
the $k$ $SU(m)$ factors are simple SQCD models and are related under
Seiberg's duality \seiberg.

\subsec{A New $D_{k+2}$ Duality: $SO(N_c)$ with Two Symmetric Tensors}

We consider a theory with $SO(N_c)$ gauge group and the following matter 
content:
\thicksize=1pt
\vskip12pt
\begintable
\tstrut  |$SO(\ncl)$  |$SU(\nfl)$|$U(1)_R$ \crthick
$\Ql$| ${\bf\ncl}$ | ${\bf \nfl}$ | $1-{\ncl-4k-2\over\nfl (k+1)}$   \cr
$X$ |  ${\bf{\ncl(\ncl+1)\over 2} - 1}$ | ${\bf 1}$ |  ${2\over k+1}$    \cr
$Y$ |  ${\bf{\ncl(\ncl+1)\over 2} - 1}$ | ${\bf 1}$ |  ${k\over k+1}$
\endtable
\noindent
The superpotential for the theory is
\eqn\soDkpot{
W={\Tr X^{k+1}\over k+1}+\Tr XY^2+\lambda_1\Tr X 
+ \lambda_2 \Tr Y } where $k$ is odd.  
The vacua of the theory are given by
\eqn\soDkeq{
X^k+Y^2 + \lambda_1 =0 = XY+YX + \lambda_2 \ .}
As in the previous case, these
equations truncate the chiral ring of the theory.  Ignoring
the Lagrange multipliers, we have
\eqn\truncA{
0={1\over2}(YX^k+X^kY) + Y^3 = Y^3}
so the meson composites
$(M_{jl}) = QX^jY^lQ$ are independent operators only for 
$j<k$ and $l<3$. The transformation properties of the $M_{jl}$ composites 
under the flavor symmetry may be determined by considering the $rs$ 
element of the $N_f\times N_f$ matrix $M_{jl}$.
\eqn\mestrans{(M_{jl})^{rs} = Q^rX^jY^lQ^s = (-1)^{jl}Q^rY^lX^jQ^s}
where we have used the vacuum condition \soDkeq\ to anticommute $Y$ through 
$X$, picking up a minus sign each time we do so.
Now, we can use the fact that the transpose of a gauge singlet is itself
to show that 
\eqn\mestrunc{\eqalign{(-1)^{jl}Q^rY^lX^jQ^s = & (-1)^{jl}(Q^rY^lX^jQ^s)^T 
= (-1)^{jl}Q^s(X^T)^j(Y^T)^lQ^r \cr 
= & (-1)^{jl}Q^sX^jY^lQ^r = (-1)^{jl}(M_{jl})^{sr}}}
where we have used the fact that $Y$ and $X$ are symmetric.
We see that for $jl$ even the meson composite $M_{jl}$ is in the symmetric 
representation whereas for $jl$ odd the meson is in the 
anti-symmetric representation of $SU(N_f)$.

The magnetic dual of this theory is an $SO(3kN_f + 8k + 4 - N_c)$ gauge 
group with charged matter 
\thicksize=1pt
\vskip12pt
\begintable
\tstrut  |$SO(\ncl)$  |$SU(\nfl)$|$U(1)_R$ \crthick
$\ql$| ${\bf\ncld}$ | ${\bf \overline\nfl}$ |$1-{\ncld-4k-2\over\nfl (k+1)}$   
\cr
$\bar X$ | ${\bf{\ncld(\ncld+1)\over 2} - 1}$ | ${\bf 1}$ |  ${2\over k+1}$    
\cr
$\bar Y$ | ${\bf{\ncld(\ncld+1)\over 2} - 1}$ | ${\bf 1}$ |  ${k\over k+1}$
\endtable
\noindent
and with gauge singlets $(M_{jl})^{rs}$
which are the images of the composites $Q^rX^jY^lQ^s$ in the electric
theory.  The superpotential is
\eqn\soDkmag{
W={\Tr \bar X^{k+1}\over k+1}+\Tr \bar X\bar Y^2 +
\sum_{j=0}^{k-1}\sum_{l=0}^2 
M_{jl}q\bar X^{k-j-1}\bar Y^{2-l} q +\lambda_1 \Tr \bar X 
+ \lambda_2 \Tr \bar Y \ .}

As evidence for this claim, we cite the following.
\item{1.}
The flavor symmetries of the two theories have the same global anomalies.
\item{2.}
Large classes of non-redundant chiral operators appear in both
theories.\foot{As in Refs.~\refs{\kenisosp,\rlmssosp,\ilslist} there are
unresolved issues involving certain baryon operators for $SO$ groups.}
\item{3.}
For $k=1$, the fields $X$ and $\bar X$ are massive; when integrated
out they leave quartic superpotentials for $Y$ and $\bar Y$.  These
theories are then dual under a duality already known from
Ref.~\kenisosp.
\item{4.}  These theories have flat directions, analogous to those of 
Eq.~\vevxy, under which they flow to a known duality.    As one
moves out along these flat direction the electric theory breaks to
$SO(n)^2\times SO(m)^k$. The symmetric tensor field $X$ contributes a
field $F$ charged as $({\bf n,n})$ under the $SO(n)\times SO(n)$
group, with superpotential $W = \Tr F^{k+1}$.  The dual theory
breaks to $SO([k+1](N_f+2) - N_f - n)^2 \times SO(N_f + 4 - m)^k$ with
similar matter content.  These first two factors are an $A_k$ loge
model, related as
described in Sec.~2.10 of \ilslist, while the latter $k$ factors are dual as
discussed in \refs{\seiberg,\kinsso}.

The most significant difference between this model and the $SU(N_c)$
model of the previous section is in the flavor representations of the
mesons.  As noted in \mestrans\ and \mestrunc, 
the mesons of this model are in the
symmetric and anti-symmetric representations of the $SU(N_f)$ flavor
group.  The mesons $M_{j1}$, in particular, are symmetric for even $j$
and anti-symmetric for odd $j$, and since $k$ is odd there are
${k+1\over 2}$ symmetric $M_{j1}$ and ${k-1\over 2}$ anti-symmetric
$M_{j1}$.  It is therefore very important for the anomaly matching
that $k$ is odd.  If one naively attempts to continue $k$ to even
values, one finds that the global anomalies cannot be made to match, because
the number of symmetric and anti-symmetric $M_{j1}$ would now be
equal.  This is unlike the $SU(N_c)$ $D_{k+2}$ model, where all mesons
are in the $({\bf N_f,\bar{N_f}})$ representation of its $SU(N_f)\times
SU(N_f)$ flavor symmetry, allowing a natural anomaly matching for
even $k$ \brodie.  Is this naive matching a mere accident?  It is
certainly disturbing, given that the chiral ring does not truncate
classically for even $k$, and there is no clear reason to restrict $j<k$ and 
$l<3$.  This issue remains unresolved at the time of writing.

\subsec{Another 
$D_{k+2}$ Orchestra Model: $Sp(N_c)$ with Two Anti-symmetric Tensors}

We consider a theory of $Sp(N_c)$ gauge group with matter content
\thicksize=1pt
\vskip12pt
\begintable
\tstrut  |$Sp(\ncl)$  |$SU(\nfl)$|$U(1)_R$ \crthick
$\Ql$| ${\bf2\ncl}$ | ${\bf 2\nfl}$ | $1-{2\ncl+4k+2\over2\nfl (k+1)}$ \cr
$X$ |  ${\bf{2\ncl(2\ncl-1)\over 2}-1}$ | ${\bf 1}$ |  ${2\over k+1}$    \cr
$Y$ |  ${\bf{2\ncl(2\ncl-1)\over 2}-1}$ | ${\bf 1}$ |  ${k\over k+1}$
\endtable
\noindent
The superpotential for the theory is of the $D_{k+2}$ form \soDkpot\ where 
$k$ is odd. 
The operators $Q^rX^jY^lQ^s$ satisfy
\eqn\truncC{
Q^rX^jY^lQ^s = (-1)^{jl}Q^rY^lX^jQ^s 
= (-1)^{jl}(Q^rY^lX^jQ^s)^T =(-1)^{jl+1}Q^sX^jY^lQ^r } and so
are in symmetric (anti-symmetric) representations of the 
flavor symmetry for $jl$ odd (even).

The magnetic dual of this theory is an $Sp(3kN_f - 4k - 2 - N_c)$ gauge
group with charged matter
\thicksize=1pt
\vskip12pt
\begintable
\tstrut  |$Sp(\ncld)$  |$SU(\nfl)$|$U(1)_R$ \crthick
$\ql$| ${\bf2\ncld}$ | ${\bf 2\overline\nfl}$ |$1-{2\ncld+4k+2\over 2\nfl 
(k+1)}$   \cr
$\bar X$ |  ${\bf{2\ncld(2\ncld-1)\over 2}-1}$ | ${\bf 1}$ |  ${2\over k+1}$
\cr
$\bar Y$ |  ${\bf{2\ncld(2\ncld-1)\over 2}-1}$ | ${\bf 1}$ |  ${k\over k+1}$
\endtable
\noindent
and with gauge singlets $(M_{jl})^{rs}$
which are the images of the composites $Q^rX^jY^lQ^s$ in the electric
theory.  The superpotential has the same form as \soDkmag.
As in the previous cases, the usual consistency checks on the duality, such
as the 't Hooft anomaly matching conditions, non-redundant operators, and 
the correspondence of $k=1$ with known dualities, are satisfied.

When $N_c$ is of the form $2n+km$ where $n$ and $m$ are integers,
there exist flat directions in the theory analogous to the ones discussed
in Sec.~4.1.  As one moves out along these flat direction the
electric theory breaks to $Sp(n)^2\times Sp(m)^k$. The anti-symmetric
tensor field $X$ contributes a field $F$ charged as $({\bf n,n})$
under the $Sp(n)\times Sp(n)$ group, with superpotential $W = \Tr
F^{k+1}$.  The dual theory breaks to $Sp((k+1)(N_f-1) - N_f - n)^2
\times Sp(N_f - 2 - m)^k$ with similar matter content. The
$Sp(n)\times Sp(n)$ factor has the $A_k$ loge duality discussed in
Sec.~2.9 of \ilslist, while the duality discussed in
\refs{\seiberg,\kippsp} holds in the $Sp(m)$ factors.

\newsec{The Stage Models}

\subsec{$SU(N_c)\times SU(N_c')$}

This theory has a gauge group $SU(N_c) \times SU(N_c')$ with matter 
content
\thicksize=1pt
\vskip12pt
\begintable
\tstrut  | $SU(\ncl)$ | $SU(\ncr)$ | $SU(\nfl)_L$ | $SU(\nfr)_L$ | 
$SU(\nfl)_R$
 | $SU(\nfr)_R$ | $U(1)_R$ \crthick
$\Ql;\Qlt$| ${\bf\ncl;\overline\ncl}$ |${\bf 1;1}$ | 
 ${\bf \nfl;1}$ | ${\bf 1;1}$ |   ${\bf 1;\nfl}$|  ${\bf 1;1}$ 
| $1+{\ncr - 2\ncl\over\nfl (k+1)}$   \cr
$\Qr;\Qrt$|${\bf 1;1}$| ${\bf\ncr;\overline\ncr}$ |${\bf 1;1}$ | 
 ${\bf \nfr;1}$ | ${\bf 1;1}$ | ${\bf 1;\nfr}$
| $1+{\ncl - 2\ncr\over\nfr (k+1)}$   \cr
$F;\tilde F$|${\bf\ncl;\overline\ncl}$| ${\bf\ncr;\overline\ncr}$ 
|${\bf 1;1}$ |${\bf 1;1}$ | ${\bf 1;1}$ | ${\bf 1;1}$
| ${k\over k+1}$   \cr
$X_1$ | ${\bf\ncl^2 -1}$ | ${\bf 1}$ | ${\bf 1}$ | ${\bf 1}$| ${\bf 1}$
|  ${\bf 1}$ |  ${2\over k+1}$    \cr
$X_2$ |${\bf 1}$|${\bf\ncr^2 -1}$ | ${\bf 1}$ | ${\bf 1}$ | ${\bf 1}$ | 
${\bf 1}$ |   ${2\over k+1}$
\endtable
\noindent
The superpotential is 
\eqn\AsupND{W = \Tr X_1^{k+1}
+ \Tr X_2^{k+1}
+ \Tr X_1\tilde F F - \Tr X_2\tilde F F + \lambda_1\Tr  X_1
+ \lambda_2 \Tr X_2. }
Here $k$ can be any positive integer.
It follows from the conditions for a supersymmetric vacuum
that the chiral ring truncates.
The gauge invariant mesons in the theory are 
$QX_1^j\tilde Q$, 
$Q'X_2^j\tilde Q'$,
$QX_1^j\tilde F Q'$,
$\tilde Q F X_2^j\tilde Q'$,
$Q\tilde F F X_1^j\tilde Q$,
$Q'\tilde FFX_2^j\tilde Q'$,
where $j=0\cdots k-1$.

The dual theory is described by an
$SU(2kN_f' + kN_f - N_c') \times SU(2kN_f + kN_f' - N_c)$ gauge theory with
matter content
\thicksize=1pt
\vskip12pt
\begintable
\tstrut  | $SU(\ncld)$ | $SU(\ncrd)$ | $SU(\nfl)_L$ | $SU(\nfr)_L$ 
 | $SU(\nfl)_R$ | $SU(\nfr)_R$ | $U(1)_R$ \crthick
$\ql;\qlt$| ${\bf\ncld;\overline\ncld}$ |${\bf 1;1}$ |  ${\bf 1;1}$ |
 ${\bf \overline\nfr;1}$ | ${\bf 1;1}$ |   ${\bf 1;\overline\nfr}$ 
| $1+{\ncrd - 2\ncld\over\nfr (k+1)}$   \cr
$\qr;\qrt$|${\bf 1;1}$| ${\bf\ncrd;\overline\ncrd}$ | 
 ${\bf \overline\nfl;1}$ | ${\bf 1;1}$ | ${\bf 1;\overline\nfl}$| ${\bf 1;1}$ 
| $1+{\ncld - 2\ncrd\over\nfl (k+1)}$   \cr
$\bar F;\tilde{\bar F}$|${\bf\ncld;\overline\ncld}$
| ${\bf\ncrd;\overline\ncrd}$ 
|${\bf 1;1}$ |${\bf 1;1}$ | ${\bf 1;1}$ | ${\bf 1;1}$
| ${k\over k+1}$   \cr
$\bar X_1$ | ${\bf\ncld^2 -1}$ | ${\bf 1}$ | ${\bf 1}$ | ${\bf 1}$| ${\bf 1}$
|  ${\bf 1}$ |  ${2\over k+1}$    \cr
$\bar X_2$ |${\bf 1}$|${\bf(\ncrd)^2 -1}$ | ${\bf 1}$ | ${\bf 1}$ 
| ${\bf 1}$ | ${\bf 1}$ |   ${2\over k+1}$
\endtable
\noindent
There are also gauge singlet mesons in the magnetic theory which
are the images of mesons in the electric theory \brodie. The dual
superpotential is analogous to \AsupND\ with the addition of coupling terms
between singlets and dual mesons.

\subsec{RG Flow from $D_{k+2}$ Orchestra Model to Stage Model }

In this section, we will discuss a deformation of the $D_{k+2}$ model
of Sec.~4.1 under which it flows to the stage model discussed above.  In
\brodie\ the construction of the stage model was motivated
by looking at the $D_{k+2}$ model with $k$ even, for which the duality
is not clear.  Here, we will see that we can in fact rigorously
derive it from the well-understood duality of the $k$-odd case.

Consider the superpotential deformed by even powered operators
\eqn\AdefW{W = \sum_{r=1}^{k+1\over 2} {s_r\over 2r} \Tr X^{2r} + \Tr XY^2
 + \lambda_1 \Tr X + \lambda_2 \Tr Y}	
where, as usual, we have taken $k$ to be odd.
The conditions for a supersymmetric minimum are
\eqn\eomeven{\eqalign{\sum_{r=1}^{k-1\over 2} s_r X^{2r} + Y^2
- \lambda_1 = 0\cr
XY+YX -\lambda_2 = 0. \cr}}
We can consider tuning the couplings $s_r$ such these become
\eqn\Aeomtune{\eqalign{X(X-a)^{k-1\over 2}(X+a)^{k-1\over 2} + Y^2
- \lambda_1= 0 \cr
XY + YX - \lambda_2 = 0. \cr}}
One solution to these equations and
 ${dW\over d\lambda_1} = 0$ has $\vev{\lambda_1}=\vev{Y}=0$ and 
\eqn\vevx{\vev{X}=\left(\matrix{a_n&0&0\cr
                  0&-a_n&0\cr
                  0&0&0_m\cr}\right)}
where a subscript $n$ indicates that an element is proportional to the 
$n\times n$ dimensional unit matrix.
The gauge group breaks from $SU(N_c)$ to $SU(n)\times SU(n)\times SU(m)
\times U(1)^2$. The off-diagonal components of $X$ are eaten in the Higgs
mechanism leaving adjoint fields $X_1$, $X_2$, and $X_3$ charged under
their respective gauge groups, of which the last is massive. The field
$Y$ breaks into massless fields $F$ and $\tilde F$ charged under
$SU(n)\times SU(n)$, a massless adjoint field $Y_3$ charged under
$SU(m)$, and some massive components.   The low-energy superpotential is
\eqn\And{W = \Tr X_1^{k+1\over 2} + \Tr X_2^{k+1\over 2} + \Tr X_1 \tilde F F
 - \Tr X_2 \tilde F F - {\Tr Y_3^4\over 2s_1} +{(\Tr Y_3^2)^2\over 2s_1m}.}  
The dual gauge group breaks similarly from $SU(3kN_f - N_c)$ to
$SU({3(k-1)\over 2}N_f - n)
\times SU({3(k-1)\over 2}N_f - n)\times SU(3N_f - m)\times U(1)^2.$
The stage model duality relates the $SU(n)\times SU(n)$ factor to the
$SU({3(k-1)\over 2}N_f - n)
\times SU({3(k-1)\over 2}N_f - n)$ factor, 
while the $SU(m)$ theory and its dual are of $A_3$ type \kutsch.

\subsec{A New Stage Model: $SO(N_c)\times SO(N_c')$}

The electric $SO(N_c)\times SO(N_c')$ theory has matter content
\thicksize=1pt
\vskip12pt
\begintable
\tstrut  | $SO(\ncl)$ | $SO(\ncr)$ | $SU(\nfl)$ | $SU(\nfr)$ | $U(1)_R$ 
\crthick
$\Ql$| ${\bf\ncl}$ |${\bf 1}$ | 
 ${\bf \nfl}$ | ${\bf 1}$ | $1+{\ncr - 2\ncl + 4k\over\nfl (k+1)}$   \cr
$\Qr$|${\bf 1}$| ${\bf\ncr}$ |${\bf 1}$ | 
 ${\bf \nfr}$ | $1+{\ncl - 2\ncr + 4k\over\nfr (k+1)}$   \cr
$F$|${\bf\ncl}$| ${\bf\ncr}$ 
|${\bf 1}$ | ${\bf 1}$
| ${k\over k+1}$   \cr
$X_1$ | ${\bf{\ncl(\ncl+1)\over 2}-1}$ | ${\bf 1}$ | ${\bf 1}$
|  ${\bf 1}$ |  ${2\over k+1}$    \cr
$X_2$ |${\bf 1}$|${\bf{\ncr(\ncr+1)\over 2}-1}$ | ${\bf 1}$ | 
${\bf 1}$ |   ${2\over k+1}$
\endtable
\noindent
The electric superpotential is
\eqn\soAkAkpot
{W=\Tr X_1^{k+1}+\Tr X_2^{k+1}+\Tr X_1F^2-\Tr X_2F^2 
+\lambda_1\Tr X_1 +\lambda_2\Tr X_2\ .}
This superpotential truncates the chiral ring for all values of
$k$. The mesons $M_j = QX_1^jQ$, $M'_j = Q'X_2^jQ'$, $N_j =
QX_1^jF^2Q$, and $N'_j = Q'X_2^jF^2Q'$ are symmetric under the
flavor groups, while the mesons $P_j = QX_1^jFQ'$ are in the
$({\bf N_f,N_f'})$ representation.

The magnetic theory is of similar type, with gauge group 
$SO(2kN_f' + kN_f + 8k - N_c')\times SO(2kN_f + kN_f' + 8k - N_c)$
and matter content
\thicksize=1pt
\vskip12pt
\begintable
\tstrut  | $SO(\ncld)$ | $SO(\ncrd)$ | $SU(\nfl)$ | $SU(\nfr)$ 
| $U(1)_R$ \crthick
$\ql$| ${\bf\ncld}$ |${\bf 1}$ |  ${\bf 1}$ | ${\bf\overline\nfr}$ 
| $1+{\ncrd - 2\ncld + 4k\over\nfr (k+1)}$   \cr
$\qr$|${\bf 1}$| ${\bf\ncrd}$ | 
 ${\bf \overline\nfl}$ | ${\bf 1}$ 
| $1+{\ncld - 2\ncrd + 4k\over\nfl (k+1)}$   \cr
$\bar F$|${\bf\ncld}$| ${\bf\ncrd}$ 
|${\bf 1}$ | ${\bf 1}$
| ${k\over k+1}$   \cr
$\bar X_1$ | ${\bf{\ncld(\ncld+1)\over 2}-1}$ | ${\bf 1}$ | ${\bf 1}$
|  ${\bf 1}$ |  ${2\over k+1}$    \cr
$\bar X_2$ |${\bf 1}$|${\bf{(\ncrd(\ncrd+1)\over 2}-1}$ | ${\bf 1}$ 
| ${\bf 1}$ |   ${2\over k+1}$
\endtable
\noindent
The mesons $M,M',N,N',P$ of the electric theory are mapped to singlets
of the dual theory in the usual way.  The dual superpotential has the
form
\eqn\BsupND{\eqalign{W &= \Tr \bar X_1^{k+1}
+  \Tr \bar X_2^{k+1}
+ \Tr \bar X_1\bar F^2
- \Tr \bar X_2\bar F^2 + \lambda_1 \Tr \bar X_1
+ \lambda_2 \Tr \bar X_2 \cr 
& +  \sum_{j=0}^{k-1} \Bigg\{
    M_j q' \bar X_2^{k-j-1} \bar F^2 q'
+   M'_j q  \bar X_1^{k-j-1} \bar F^2 q \cr
 & \qquad\qquad\qquad +  P_jq  \bar X_1^{k-j-1} \bar F q'
+   N_j q' \bar X_2^{k-j-1} q'
+   N'_j q  \bar X_1^{k-j-1} q\Bigg\} \ . \cr}}

\subsec{RG Flow from $D_{k+2}$ Orchestra Model to Stage Model}  

In analogy to Sec.~5.2, deforming the superpotential of the
$SO(N_c)$ gauge theory discussed in Sec.~4.2 by even powers $\Tr
X^{2r}$ can cause it to flow from the $D_{k+2}$ orchestra theory to
this stage model.  The superpotential has
the form
\eqn\BdefW{W = \sum_{r=1}^{k+1\over 2} {s_r\over 2r} \Tr X^{2r} + \Tr XY^2
 - \lambda_1 \Tr X - \lambda_2 \Tr Y}
where, as usual, we have taken $k$ to be odd.
As described in Sec.~5.2, by tuning the couplings $s_r$ we can find a
superpotential which has a vacuum $<Y>=0$ and
\eqn\vevxA{<X>=\left(\matrix{a_n&0&0\cr
                  0&-a_n&0\cr
                  0&0&0_m\cr}\right)}
The electric theory breaks from $SO(N_c)$ down to $SO(n)\times SO(n)
\times SO(m)$.
The low energy superpotential is
\eqn\Bnd{W = \Tr X_1^{k+1\over 2} + \Tr X_1 F^2
+ \Tr X_2^{k+1\over 2}
- \Tr X_2 F^2 - {\Tr Y_3^4\over 2s_1} + {(\Tr Y_3^2)^2\over 2s_1m}.}
where the fields $X_1$, $X_2$, are symmetric tensors coming from $X$,
$Y_3$ is a symmetric tensor and $F$ is a field in the $({\bf n,n})$
representation coming from $Y$. The dual
gauge group breaks from $SO(3kN_f + 8k + 4 - N_c)$ to $SO(3{(k-1)\over
2}N_f + 4k - 4 - n)\times SO(3{(k-1)\over 2}N_f + 4k - 4 - n) 
\times SO(3N_f + 12 - m)$, which is dual to the low-energy electric
theory under the stage and $A_3$
\kenisosp\ duality transformations.

\subsec{Another New Stage Model: $Sp(N_c)\times Sp(N_c')$}

Consider the gauge theory $Sp(N_c)\times Sp(N_c')$ with matter content
\thicksize=1pt
\vskip12pt
\begintable
\tstrut  | $Sp(\ncl)$ | $Sp(\ncr)$ | $SU(\nfl)$ | $SU(\nfr)$ | $U(1)_R$ 
\crthick
$\Ql$| ${\bf2\ncl}$ |${\bf 1}$ | 
 ${\bf 2\nfl}$ | ${\bf 1}$ | $1+{\ncr - 2\ncl - 2k\over\nfl (k+1)}$   \cr
$\Qr$|${\bf 1}$| ${\bf2\ncr}$ |${\bf 1}$ | 
 ${\bf 2\nfr}$ | $1+{\ncl - 2\ncr - 2k\over\nfr (k+1)}$   \cr
$F$|${\bf2\ncl}$| ${\bf2\ncr}$ 
|${\bf 1}$ | ${\bf 1}$
| ${k\over k+1}$   \cr
$X_1$ | ${\bf{2\ncl(2\ncl-1)\over 2}-1}$ | ${\bf 1}$ | ${\bf 1}$
|  ${\bf 1}$ |  ${2\over k+1}$    \cr
$X_2$ |${\bf 1}$|${\bf{2\ncr(2\ncr-1)\over 2}-1}$ | ${\bf 1}$ | 
${\bf 1}$ |   ${2\over k+1}$
\endtable
\noindent
The electric superpotential takes the form \soAkAkpot\ and 
truncates the chiral ring for all values of $k$. The 
mesons in the theory anti-symmetric under the flavor groups are 
$QX_1^jQ$, $Q'X_2^jQ'$, $QX_1^jF^2Q$, and $Q'X_2^jF^2Q'$. 
The mesons in the theory in the $({\bf 2N_f,2N_f'})$ representation are
$QX_1^jFQ'$.

The magnetic theory is of similar type, with gauge group 
$Sp(2kN_f' + kN_f - 4k - N_c')\times Sp(2kN_f + kN_f' - 4k - N_c)$
and matter content
\thicksize=1pt
\vskip12pt
\begintable
\tstrut  | $Sp(\ncld)$ | $Sp(\ncrd)$ | $SU(\nfl)$ | $SU(\nfr)$ 
| $U(1)_R$ \crthick
$\ql$| ${\bf2\ncld}$ |${\bf 1}$ |  ${\bf 1}$ | ${\bf 2\overline\nfr}$ 
| $1+{\ncrd - 2\ncld - 2k\over\nfr (k+1)}$   \cr
$\qr$|${\bf 1}$| ${\bf2\ncrd}$ | 
 ${\bf 2\overline\nfl}$ | ${\bf 1}$ 
| $1+{\ncld - 2\ncrd - 2k\over\nfl (k+1)}$   \cr
$\bar F$|${\bf2\ncld}$| ${\bf2\ncrd}$ 
|${\bf 1}$ | ${\bf 1}$
| ${k\over k+1}$   \cr
$\bar X_1$ | ${\bf{2\ncld(2\ncld-1)\over 2}-1}$ | ${\bf 1}$ | ${\bf 1}$
|  ${\bf 1}$ |  ${2\over k+1}$    \cr
$\bar X_2$ |${\bf 1}$|${\bf{2\ncrd(2\ncrd-1)\over 2}-1}$ | ${\bf 1}$ | ${\bf 1}$ 
|   ${2\over k+1}$
\endtable
\noindent
The mesons of the electric theory are mapped to singlets of the dual theory
in the usual way.

In analogy to Sec.~5.2, deforming the $Sp(N_c)$ $D_{k+2}$
superpotential by even powers $\Tr X^{2r}$ can cause the $Sp(N_c)$
orchestra theory of Sec.~4.3 to flow to the stage model of this
section. When the electric theory breaks to $Sp(n)\times Sp(n)\times
Sp(m)$, the dual gauge group breaks from $Sp(3kN_f - 4k - 2 - N_c)$ to
$Sp(3{(k-1)\over 2}N_f - 2k + 2 - n)
\times Sp(3{(k-1)\over 2}N_f - 2k + 2 - n) \times Sp(3N_f - 6 - m)$.
The $Sp(n)\times Sp(n)$ factors are dual as described above, and
the $A_3$ duality of \kenisosp\ applies in the $Sp(m)$ sector.

\newsec{$D_{k+2}$ Mezzanine Dualities}

\subsec{Mezzanine: $SO(N_c)$ with a Symmetric and an Anti-symmetric Tensor}

We consider an $SO(N_c)$ gauge theory with 
\thicksize=1pt
\vskip12pt
\begintable
\tstrut  |$SO(\ncl)$  |$SU(\nfl)$|$U(1)_R$ \crthick
$\Ql$| ${\bf\ncl}$ | ${\bf \nfl}$ | $1-{\ncl-4k+2\over\nfl (k+1)}$   \cr
$X$ |  ${\bf{\ncl(\ncl+1)\over 2}-1}$ | ${\bf 1}$ |  ${2\over k+1}$    \cr
$Y$ |  ${\bf\ncl(\ncl-1)\over 2}$ | ${\bf 1}$ |  ${k\over k+1}$
\endtable
\noindent
The superpotential for the theory is of the $D_{k+2}$ form 
\eqn\mezDkpot{
W={\Tr X^{k+1}\over k+1}+\Tr XY^2 + \lambda \Tr X \ .} 
$k$ is odd.  
The meson composites
$QX^jY^lQ$ are independent operators only for $j<k$ and $l<3$.
Furthermore, the operators $Q^rX^jY^lQ^s$ satisfy
\eqn\truncD{
Q^rX^jY^lQ^s = (-1)^{jl}Q^rY^lX^jQ^s 
= (-1)^{jl}(Q^rY^lX^jQ^s)^T =(-1)^{(j+1)l}Q^sX^jY^lQ^r } and so
are in symmetric (anti-symmetric) representations of the
flavor symmetry for $(j+1)l$ even (odd).

The magnetic dual of this theory is an $SO(3kN_f + 8k - 4 - N_c)$ gauge 
group with charged fields
\thicksize=1pt
\vskip12pt
\begintable
\tstrut  |$SO(\ncld)$  |$SU(\nfl)$|$U(1)_R$ \crthick
$\ql$| ${\bf\ncld}$ | ${\bf \overline\nfl}$ |$1-{\ncld-4k+2\over\nfl (k+1)}$   
\cr
$\bar X$ |  ${\bf{\ncld(\ncld+1)\over 2}-1}$ | ${\bf 1}$ |  ${2\over k+1}$    
\cr
$\bar Y$ |  ${\bf\ncld(\ncld-1)\over 2}$ | ${\bf 1}$ |  ${k\over k+1}$
\endtable
\noindent
and with gauge singlets $(M_{jl})^{rs}$
which are the images of the composites $Q^rX^jY^lQ^s$ in the electric
theory.  The dual superpotential is of the form 
\eqn\mezDkmag{
W={\Tr \bar X^{k+1}\over k+1}+\Tr \bar X\bar Y^2 
+ \sum_{j=0}^{k-1}\sum_{l=0}^2 
M_{jl}q\bar X^{k-j-1}\bar Y^{2-l} q +\lambda \Tr \bar X \ .}
As in the previous cases, we find that the usual consistency checks on
the duality are satisfied.

In analogy to Sec.~5.2, deformations of the superpotential by even
powers of $\Tr X^{2r}$ can cause the theory to flow to the stage model 
discussed in Sec.~5.3.  When the electric theory breaks
to $SO(n)\times SO(n)\times SO(m)$, the dual gauge group breaks from
$SO(3kN_f + 8k - 4 - N_c)$ to $SO(3{(k-1)\over 2}N_f + 4k - 4 - n)
\times SO(3{(k-1)\over 2}N_f + 4k - 4 - n) \times SO(3N_f + 4 - m)$. 
The duality of Sec.~5.3 applies in the $SO(n)\times SO(n)$ sector
and the duality of \rlmssosp\ applies in the $SO(m)$ sector.

\subsec{Mezzanine: $Sp(N_c)$ with an Anti-symmetric and a Symmetric Tensor}

We consider a theory with $Sp(N_c)$ gauge group and matter content
\thicksize=1pt
\vskip12pt 
\begintable
\tstrut  |$Sp(\ncl)$  |$SU(\nfl)$|$U(1)_R$ \crthick
$\Ql$| ${\bf2\ncl}$ | ${\bf 2\nfl}$ | $1-{\ncl+2k-1\over\nfl (k+1)}$   \cr
$X$ |  ${\bf{2\ncl(2\ncl-1)\over 2}-1}$ | ${\bf 1}$ |  ${2\over k+1}$    \cr
$Y$ |  ${\bf{2\ncl(2\ncl+1)}\over 2}$ | ${\bf 1}$ |  ${k\over k+1}$
\endtable
\noindent
The superpotential for the theory is of the $D_{k+2}$ form 
\mezDkpot\ where $k$ is odd. 
Furthermore, the operators $Q^rX^jY^lQ^s$ satisfy
\eqn\truncE{
Q^rX^jY^lQ^s = (-1)^{jl}Q^rY^lX^jQ^s
= (-1)^{jl}(Q^rY^lX^jQ^s)^T =(-1)^{jl+l+1}Q^sX^jY^lQ^r } and so
are in symmetric (anti-symmetric) representations of the 
flavor symmetry for $(j+1)l$ odd (even).

The magnetic dual of this theory is an $Sp(3kN_f - 4k + 2 - N_c)$ gauge
group with charged matter
\thicksize=1pt
\vskip12pt
\begintable
\tstrut  |$Sp(\ncld)$  |$SU(\nfl)$|$U(1)_R$ \crthick
$\ql$| ${\bf2\ncld}$ | ${\bf 2\overline\nfl}$ 
|$1-{\ncld+2k-1\over\nfl (k+1)}$   \cr
$\bar X$ |  ${\bf{2\ncld(2\ncld-1)\over 2}-1}$ | ${\bf 1}$ |  ${2\over k+1}$    
\cr
$\bar Y$ |  ${\bf{2\ncld(2\ncld+1)\over 2}}$ | ${\bf 1}$ |  ${k\over k+1}$
\endtable
\noindent
and with gauge singlets $(M_{jl})^{rs}$
which are the images of the composites $Q^rX^jY^lQ^s$ in the electric
theory.  The superpotential has the form \mezDkmag.
All the usual consistency checks are satisfied.

As in Sec.~5.2, by deforming the superpotential by even powers $\Tr
X^{2r}$, one can flow from this mezzanine theory to the stage 
model discussed in Sec.~5.5.  When the electric theory breaks to
$Sp(n)\times Sp(n)\times Sp(m)$, the dual gauge group breaks from
$Sp(3kN_f - 4k + 2 - N_c)$ to $Sp(3{(k-1)\over 2}N_f - 2k + 2 - n)
\times Sp(3{(k-1)\over 2}N_f - 2k + 2 - n) \times Sp(3N_f - 2 - m).$
The duality in the $Sp(n)\times Sp(n)$ sector is that of Sec.~5.5 
and the duality in the $Sp(m)$ sector is that of \rlmssosp.

\subsec{Mezzanine: $SU(N_c)$ with  Adjoint, Symmetric, 
and Conjugate Symmetric Tensors}

The $SU(N_c)$ gauge theory has the following matter content:
\thicksize=1pt
\vskip12pt
\begintable
\tstrut  |$SU(\ncl)$  |$SU(\nfl)_L$|$SU(\nfl)_R$|$U(1)_Y$|$U(1)_B$|$U(1)_R$ 
\crthick
$\Ql$| ${\bf\ncl}$ |${\bf\nfl}$ | 
 ${\bf 1}$ | 0  | ${1\over N_c}$ | $1-{\ncl -2\over\nfl (k+1)}$   \cr
$\Qlt$| ${\bf\overline\ncl}$ |${\bf 1}$ | 
 ${\bf \overline\nfl}$ | 0  | $-{1\over N_c}$ | $1-{\ncl -2\over\nfl (k+1)}$   
\cr
$X$ | ${\bf\ncl^2 -1}$ | ${\bf 1}$ | ${\bf 1}$ | 0 | 0 |  ${2\over k+1}$  \cr
$Y$ | ${\bf\ncl(\ncl+1)\over 2}$ | ${\bf 1}$ | ${\bf 1}$ | 1 |${2\over N_c}$ 
|${k\over k+1}$ \cr
$\tilde Y$ | $\overline{{\bf\ncl(\ncl+1)\over 2}}$ 
| ${\bf 1}$ |${\bf 1}$| -1 |$-{2\over N_c}$|${k\over k+1}$
\endtable
\noindent
The superpotential is 
\eqn\ABb{W = {\Tr X^{k+1}\over k+1} + \Tr XY\tilde Y + \lambda \Tr X}
where $\lambda$ is a Lagrange multiplier.  The conditions for a
supersymmetric minimum that follow from this superpotential are
\eqn\Beom{\eqalign{X^k + Y\tilde Y + \lambda = 0\cr 
{XY+YX^T\over 2} = 0 \cr  {X\tilde Y+\tilde YX^T\over 2} = 0. \cr}} 
These equations truncate the chiral
ring for odd values of $k$. As usual, we ignore
$\lambda$. We can
multiply $X^k$ by $Y$ from the right to form $X^kY = -Y\tilde YY$.
We can then take the transpose of the first of \Beom\ and multiply it
from the left by $Y$.
Combining these two equations we have
\eqn\ringg{Y( X^T )^k + X^k Y = -2Y{\tilde Y} Y.}
Now, we can use the second equation in \Beom\ to show that 
\eqn\ringgg{((-1)^k+ 1)X^k Y = -2Y{\tilde Y} Y.}
Thus for odd $k$, $Y{\tilde Y} Y=0$.
Likewise, ${\tilde Y} Y \tilde Y = 0$.
In this theory, one can construct gauge invariant mesons in the 
$({\bf N_f,\overline N_f})$ representation 
\eqn\Nmesons{\eqalign{(N_j)^{r}_s &= Q^r X^j\tilde Y Y \tilde Q_s\cr
(M_j)^{r}_s &= Q^r X^j \tilde Q_s\cr}}
 where $j = 0,1,\cdots,k-1$, while the mesons
\eqn\Pmesons{\eqalign{(P_j)^{rs} = & Q^r X^j\tilde Y Q^s = (-1)^j
Q^r \tilde Y (X^T)^j Q^s \cr = & (-1)^j(Q^r \tilde Y (X^T)^j Q^s)^T 
 = (-1)^jQ^s X^j\tilde Y Q^r
\cr
 (\tilde P_j)_{rs} = & \tilde Q_r Y X^j \tilde Q_s  = 
(-1)^j\tilde Q_r (X^T)^j Y \tilde Q_s \cr
= & (-1)^j (\tilde Q_r (X^T)^j Y \tilde Q_s)^T
= (-1)^j\tilde Q_s Y X^j \tilde Q_r
\cr }} 
are in the 
(anti)--symmetric representation of the flavor group for (odd) even 
values of $j$ where $j = 0,1, \cdots ,k-1$.

This theory has a dual description in terms of an $SU(3kN_f + 4 -
N_c)$ gauge theory with the following charged fields
\thicksize=1pt
\vskip12pt
\begintable
\tstrut  |$SU(\ncld)$  |$SU(\nfl)_L$| $SU(\nfl)_R$| $U(1)_Y$| $U(1)_B$ | 
$U(1)_R$ \crthick
$\ql$| ${\bf\ncld}$ |${\bf\overline\nfl}$ | 
 ${\bf 1}$ |${kN_f + 2\over \ncld}$ |${1\over \ncld}$ | $1-{\ncld -2\over\nfl 
(k+1)}$   \cr
$\qlt$| ${\bf\overline\ncld}$ |${\bf1}$ | 
 ${\bf \nfl}$ |$-{kN_f + 2\over \ncld}$ |$-{1\over \ncld}$ 
| $1-{\ncld -2\over\nfl (k+1)}$   \cr
$\bar X$ | ${\bf\ncld^2 -1}$ | ${\bf 1}$ | ${\bf 1}$ | 0 |0 |  ${2\over k+1}$ 
\cr
$\bar Y$|${\bf\ncld(\ncld+1)\over 2}$ | ${\bf 1}$ | ${\bf 1}$|
${N_c - kN_f\over \ncld}$ |${2\over \ncld}$ |${k\over k+1}$ 
\cr
$\tilde {\bar Y}$|$\overline{\bf\ncld(\ncld+1)\over 2}$|${\bf 1}$|${\bf 1}$
|$-{N_c - kN_f\over \ncld}$ |$-{2\over \ncld}$ |${k\over k+1}$
\endtable
\noindent
The dual description also has singlets, $(N_{j})^r_s,(M_{j})^r_s,
(P_j)^{rs},(\tilde P_j)_{rs}$, which are in
a one-to-one mapping with the mesons in Eqs.~\Nmesons-\Pmesons.  The dual
superpotential has the form
\eqn\Bf{\eqalign{
W= & {\Tr \bar X^{k+1}\over k+1}  + \Tr \bar X \bar Y\tilde {\bar Y}
+ \lambda \Tr \bar X \cr
&+ \sum_{j=0}^{k-1} \left\{N_j\tilde q \bar X^{k-j-1} q 
               + P_j       q \bar X^{k-j-1} \bar Y q 
         +\tilde P_j\tilde q \tilde {\bar Y} \bar X^{k-j-1} \tilde q 
               + M_j\tilde q \bar X^{k-j-1} \tilde {\bar Y} \bar Y q \right\}
\ . \cr}}
As in the previous cases, we find that the usual consistency checks on
the duality are satisfied.

As in Sec.~5.2, by deforming the superpotential by even powers $\Tr
X^{2r}$, one can flow from this theory to the stage model
discussed in Sec.~5.1.  When the electric group breaks to
$SU(n)\times SU(n)\times SU(m)$, the dual gauge group breaks from
$SU(3kN_f + 4 - N_c)$ to $SU({3(k-1)\over 2}N_f - n)
\times SU({3(k-1)\over 2}N_f - n)\times SU(3N_f + 4 - m)\times U(1)^2.$
The duality in the $SU(n)\times SU(n)$ part of the theory was
described in Sec.~5.1 and in \brodie; that of the $SU(m)$ part was
described in \ilslist.

\subsec{Mezzanine: $SU(N_c)$ with Adjoint, Anti-symmetric, 
and Conjugate Anti-symmetric Tensors}

The $SU(N_c)$ gauge theory has the following matter content:
\thicksize=1pt
\vskip12pt
\begintable
\tstrut  |$SU(\ncl)$  |$SU(\nfl)_L$|$SU(\nfl)_R$|$U(1)_Y$|$U(1)_B$|$U(1)_R$ 
\crthick
$\Ql$| ${\bf\ncl}$ |${\bf\nfl}$ | 
 ${\bf 1}$ | 0  | ${1\over N_c}$ |$1-{\ncl +2\over\nfl (k+1)}$   \cr
$\Qlt$| ${\bf\overline\ncl}$ |${\bf1}$ | 
 ${\bf \overline\nfl}$ | 0 | $-{1\over N_c}$
| $1-{\ncl +2\over\nfl (k+1)}$   \cr
$X$ | ${\bf\ncl^2 -1}$ | ${\bf 1}$ | ${\bf 1}$ | 0  | 0 
|  ${2\over k+1}$    \cr
$Y$ | ${\bf\ncl(\ncl-1)\over 2}$ | ${\bf 1}$ | ${\bf 1}$ | 1 
| ${2\over N_c}$ |${k\over 
k+1}$ \cr
$\tilde Y$ | $\overline{\bf\ncl(\ncl-1)\over 2}$ 
| ${\bf 1}$ |${\bf 1}$| -1 |$-{2\over N_c}$ |${k\over k+1}$
\endtable
\noindent
The superpotential is of the form \ABb, and as before 
the chiral ring truncates for $k$ odd.
In this theory, one can construct gauge invariant mesons in the 
$({\bf N_f,\overline N_f})$
representation  
\eqn\NmesonsB{\eqalign{(N_j)^{r}_s &= Q^r X^j\tilde Y Y \tilde Q_s\cr
(M_j)^{r}_s &= Q^r 
X^j \tilde Q_s\cr}}
 where $j = 0,1, \cdots ,k-1$, while the mesons 
\eqn\PmesonsB{\eqalign{P_j = & Q^r  X^j\tilde Y Q^s = (-1)^j
Q^r \tilde Y (X^T)^j Q^s \cr = & (-1)^j (Q^r \tilde Y (X^T)^j Q^s)^T 
= (-1)^{j+1} Q^s X^j\tilde Y Q^r
\cr
\tilde P_j = & \tilde Q_r Y X^j \tilde Q_s  = 
(-1)^j\tilde Q_r (X^T)^j Y \tilde Q_s \cr
= & (-1)^j(\tilde Q_r (X^T)^j Y \tilde Q_s)^T
= (-1)^{j+1} \tilde Q_s Y X^j \tilde Q_r
\cr }}
are in the 
(anti)--symmetric representation of the flavor group for (even) odd 
values of $j$ where $j = 0,1, \cdots ,k-1$.

This theory has a dual description in terms of an $SU(3kN_f - 4 - N_c)$ gauge
theory with the following charged fields:
\thicksize=1pt
\vskip12pt
\begintable
\tstrut  |$SU(\ncld)$  |$SU(\nfl)_L$|$SU(\nfl)_R$| $U(1)_Y$|  
$U(1)_B$|$U(1)_R$ \crthick
$\ql$| ${\bf\ncld}$ |${\bf\overline\nfl}$ | 
 ${\bf 1}$ |${kN_f-2\over \ncld}$ | ${1\over \ncld}$ | $1-{\ncld + 2\over\nfl 
(k+1)}$   \cr
$\qlt$| ${\bf\overline\ncld}$ |${\bf1}$ | 
 ${\bf \nfl}$ |$-{kN_f-2\over \ncld}$  |$-{1\over \ncld}$
| $1-{\ncld + 2\over\nfl (k+1)}$   \cr
$\bar X$ | ${\bf\ncld^2 -1}$ | ${\bf 1}$ | ${\bf 1}$ |0 |0 
|  ${2\over k+1}$ \cr
$\bar Y$|${\bf\ncld(\ncld-1)\over 2}$ | ${\bf 1}$ | ${\bf 1}$ 
|${N_c-kN_f\over \ncld}$ |${2\over \ncld}$ |${k\over 
k+1}$ \cr
$\tilde {\bar Y}$|$\overline{\bf\ncld(\ncld-1)\over 2}$ 
| ${\bf 1}$ |${\bf 1}$|$-{N_c-kN_f\over \ncld}$ |$-{2\over \ncld}$ 
|${k\over k+1}$
\endtable
\noindent
The dual description also has singlets, $(N_{j})^r_s,(M_{j})^r_s,
(P_j)^{rs},(\tilde P_j)_{rs}$, which are in
a one-to-one mapping with the mesons in Eqs.~\NmesonsB-\PmesonsB.  
The dual superpotential has the form \Bf.
As in the previous cases, we find that the usual consistency checks on
the duality are satisfied.

As in Sec.~5.2, by deforming the superpotential by even powers $\Tr
X^{2r}$, one can flow from this mezzanine theory to the stage model 
discussed in Sec.~5.1.  When the electric group breaks
to $SU(n)\times SU(n)\times SU(m)$, the dual gauge group breaks from
$SU(3kN_f - 4 - N_c)$ to $SU({3(k-1)\over 2}N_f - n)
\times SU({3(k-1)\over 2}N_f - n)\times SU(3N_f - 4 - m)\times U(1)^2.$
The duality of the $SU(n)\times SU(n)$ part of the theory was described 
in Sec.~5.1 and in \brodie; that of the $SU(m)$ part was described in 
Sec.~2.6 of \ilslist .

\newsec{$D_{k+2}$ Balcony Model: $SU(N_c)$ with Adjoint, Anti-symmetric
and Conjugate Symmetric Tensors}

The $SU(N_c)$ gauge theory has the following matter content:
\thicksize=1pt
\vskip12pt
\begintable
\tstrut  |$SU(\ncl)$  |$SU(m_f)$|$SU(\tilde m_f)$|$U(1)_Y$|$U(1)_B$|$U(1)_R$ 
\crthick
$\Ql$| ${\bf\ncl}$ |${\bf m_f}$ | 
 ${\bf 1}$ |${6\over m_f+4}-1$ |${1\over N_c}$ 
| $1-{\ncl + 6k\over m_f (k+1)}$   \cr
$\Qlt$| ${\bf\overline\ncl}$ |${\bf 1}$ | 
 ${\bf \tilde m_f}$ |${6\over m_f-4}+1$ |$-{1\over N_c}$ 
| $1-{\ncl - 6k\over \tilde m_f (k+1)}$   \cr
$X$ | ${\bf\ncl^2 -1}$ | ${\bf 1}$ | ${\bf 1}$ | 0 | 0 |  ${2\over k+1}$ \cr
$Y$ | ${\bf\ncl(\ncl-1)\over 2}$ | ${\bf 1}$ | ${\bf 1}$ 
    | 1 | ${2\over N_c}$ | ${k\over k+1}$ \cr
$\tilde Y$ | $\overline{\bf\ncl(\ncl+1)\over 2}$ | ${\bf 1}$ |${\bf 1}$| -1 
|$-{2\over N_c}$  |${k\over k+1}$
\endtable
\noindent
This theory is chiral, and the anomaly cancellation requirement is that
$m_f = \tilde m_f + 8$.
The superpotential is 
\eqn\ADb{W = {\Tr X^{k+1}\over k+1} + \Tr XY\tilde Y + \lambda \Tr X}
where $\lambda$ is a Lagrange multiplier. It follows from the conditions
for a supersymmetric minimum that the chiral ring truncates for all 
values of $k$, odd or even. This is in contrast with all the other examples
in this paper, for which the chiral ring truncates only for $k$ odd.
To see why the $k$ even case is different here, we consider the vacuum
conditions that follow from this superpotential.
\eqn\Ceom{\eqalign{X^k + Y\tilde Y + \lambda = 0\cr 
{XY-YX^T\over 2} = 0 \cr  {X\tilde Y-\tilde YX^T\over 2} = 0. \cr}} 
As usual, we ignore
$\lambda$. We can
multiply $X^k$ by $Y$ from the right to form $X^kY = -Y\tilde YY$.
We can then take the transpose of the first of \Ceom\ and multiply it
from the left by $Y$ to form $Y(X^T)^k = +Y\tilde YY$.
Combining these two equations we have
\eqn\ringggg{Y( X^T )^k - X^k Y = 2Y{\tilde Y} Y.}
Now, we can use the second equation in \Ceom\ to show that 
for all values of $k$, $Y{\tilde Y} Y=0$.
Likewise, ${\tilde Y} Y \tilde Y = 0$.
In this theory, one can construct gauge invariant mesons in the 
$({\bf N_f,\overline N_f})$
representation  
\eqn\NmesonsC{\eqalign{(N_j)^{r}_s &= Q^r X^j\tilde Y Y \tilde Q_s\cr
(M_j)^{r}_s &= Q^r 
X^j \tilde Q_s\cr}}
 where $j = 0,1, \cdots ,k-1$, while the mesons
\eqn\PmesonsC{\eqalign{P_j = & Q^r  X^j\tilde Y Q^s = 
Q^r \tilde Y (X^T)^j Q^s \cr = & (Q^r \tilde Y (X^T)^j Q^s)^T
 = Q^s X^j\tilde Y Q^r
\cr
\tilde P_j = & \tilde Q_r Y X^j \tilde Q_s = 
\tilde Q_r (X^T)^j Y \tilde Q_s \cr
= & (\tilde Q_r (X^T)^j Y \tilde Q_s)^T
= (-1)\tilde Q_s Y X^j \tilde Q_r
\cr }}
are in the symmetric (anti-symmetric) representation of the 
flavor group. 

This theory has a dual description in terms of a 
$SU(3k{(m_f+\tilde m_f)\over 2} - N_c)$ gauge
theory with the following matter fields
\thicksize=1pt
\vskip12pt
\begintable
\tstrut  |$SU(\ncld)$|$SU(m_f)$|$SU(\tilde m_f)$| $U(1)_Y$|$U(1)_B$|$U(1)_R$ 
\crthick
$\ql$| ${\bf\ncld}$ |${\bf\overline m_f}$ | 
 ${\bf 1}$ |$1-{6\over m_f + 4}$ | ${1\over \ncld}$ 
| $1-{\ncld + 6k\over m_f (k+1)}$   \cr
$\qlt$| ${\bf\overline\ncld}$ |${1}$ | 
 ${\bf\overline{\tilde m_f}}$ | $-1-{6\over m_f + 4}$ |$-{1\over \ncld}$ 
| $1-{\ncld - 6k\over \tilde m_f (k+1)}$   \cr
$\bar X$ | ${\bf\ncld^2 -1}$ | ${\bf 1}$ | ${\bf 1}$ |0 |0 
|  ${2\over k+1}$ \cr
$\bar Y$|${\bf\ncld(\ncld-1)\over 2}$ | ${\bf 1}$ | ${\bf 1}$ | 1 
|${2\over \ncld}$ |${k\over k+1}$ \cr
$\tilde {\bar Y}$|$\overline{\bf\ncld(\ncld+1)\over 2}$ | ${\bf 1}$ |${\bf 
1}$|-1|$-{2\over \ncld}$|${k\over k+1}$
\endtable
\noindent
The dual description also has singlets, $(N_j)^r_s,(M_j)^r_s,
(P_j)^{rs},(\tilde P_j)_{rs}$, which are in
a one-to-one mapping with the mesons in Eqs.~\NmesonsC-\PmesonsC.
The dual superpotential has the form
\eqn\Df{\eqalign{
W= & {  \Tr \bar X^{k+1}\over k+1} + \Tr \bar X \bar Y\tilde{\bar Y} 
+ \lambda \Tr \bar X \cr
& + \sum_{j=0}^{k-1} \left\{N_j\tilde q \bar X^{k-j-1} q 
               + P_j       q \bar X^{k-j-1} \bar Y q 
         +\tilde P_j\tilde q \tilde {\bar Y} \bar X^{k-j-1} \tilde q 
               + M_j\tilde q \bar X^{k-j-1} \tilde{\bar Y}\bar Y q \right\}
 \ .\cr}}
As in the previous cases, we find that the usual consistency checks on
the duality are satisfied.

For $N_c=kn$, this theory has flat directions
\eqn\vevxB{\vev{X}=a\left(\matrix{1_n&0&0&0&.&0\cr
                   0&\omega_n &0&0&.&0\cr
                   0&0&\omega_n^2&0&.&0\cr
                   0&0&0&\omega_n^3&.&0\cr
		   .&.&.&.&.&.\cr
		   0&0&0&0&.&\omega_n^{k-1}\cr}\right)}
which break the theory down to $SU(n)^k$.  In each $SU(n)$ factor, the
adjoint field from $X$ is massive while anti-symmetric and conjugate
symmetric tensors $\hat Y$ and $\hat{\tilde Y}$ are massless, with
low-energy superpotential $W = \Tr (\tilde {\hat Y} \hat Y)^2$.  The
model has broken to $k$ copies of the $A_k$ balcony model, the chiral
theory studied in \ilslist.  The dual gauge group breaks from
$SU(3k{(m_f + \tilde m_f)\over 2} - N_c)$ down to $SU(3{(m_f + \tilde
m_f)\over 2} - n)^k$, which is consistent with the duality
transformation of \ilslist.

It is also straightforward to deform the $k$-odd theory and flow to
the $k$-even theory.  Consider the deformed superpotential
\eqn\CdefW{W = \sum_{r=2}^{k+1} {s_r\over r} \Tr X^r 
+ \Tr XY\tilde Y
 - \lambda \Tr X.}
We can consider tuning the couplings $s_r$ such that the conditions
for a supersymmetric vacuum become
\eqn\Beomtune{\eqalign{(X-a)^{k-1}(X-b) + Y\tilde Y
= 0 \cr
X\tilde Y - \tilde YX^T = 0  \cr
XY - YX^T = 0. \cr}}
The tracelessness condition permits us to choose as eigenvalues of the 
$N_c\times N_c$ tensor, $X$, to be
\eqn\vevxC{\vev{X}=\left(\matrix{a_n&0\cr
                  0&b_m\cr}\right)}
where $na+mb=0$ and where a subscript $n$ indicate that the element is
proportional to the $n\times n$-dimensional unit matrix.  We also take
$\vev{Y} = \vev{\tilde Y} = 0$.  The gauge group breaks from $SU(N_c)$
to $SU(n)\times SU(m)\times U(1)$.  The off-diagonal components of $X$
are eaten in the Higgs mechanism leaving adjoint fields $X_1$ and
$X_2$ charged under their respective gauge groups.  The fields $Y$ and
$\tilde Y$ have massive fields $F$ and $\tilde F$ coming from their
off-diagonal blocks but massless anti-symmetric fields $Y_1$ and $Y_2$
and massless symmetric fields $\tilde Y_1$ and $\tilde Y_2$ coming
from their diagonal blocks. We have
\eqn\Cnd{W = \Tr X_1^k + \Tr X_1 Y_1\tilde Y_1 
+ \Tr X_2^2
+ \Tr X_2 Y_2\tilde Y_2.} 
Because of the mapping of operators, the dual gauge group breaks similarly 
from
$SU(3k{(m_f + \tilde m_f)\over 2} - N_c)$ to 
$SU(3(k-1){(m_f+\tilde m_f)\over 2} - n)
\times SU(3{(m_f+\tilde m_f)\over 2} - m).$
We have now two decoupled theories of the same type that we started with.
However, $X$ is raised to an odd power in the first vacuum 
whereas when we started it was raised to an even power.

\newsec{Baryons and Additional Checks on Mezzanine and Balcony Models}

An additional interesting check involves certain baryonic flat
directions which are present in the $D_{k+2}$ $SU(N_c)$ mezzanine and
balcony models, each of which has fields $X$, $Y$ and $\tilde Y$.  As
shown in \figAk, the $A_k$ mezzanine models have baryonic flat
directions along which they flow to $SO$ and $Sp$ orchestra models,
while the $A_k$ balcony model has baryonic flat directions along which
it can flow to the $SO$ and $Sp$ mezzanine models.  In the $D_k$
models the situation is different.  The $D_{k+2}$ $SU(N_c)$ mezzanine
models flow to the $A_k$ {\it mezzanine} models with $SO(N_c)$ and
$Sp(N_c)$ gauge group, while the $D_{k+2}$ balcony model flows to the
$A_k$ {\it orchestra} models with $SO(N_c)$ and $Sp(N_c)$ gauge group.
The consistency of the mapping of baryon operators in the $D_{k+2}$
models and their flow to known $A_k$ dualities is a significant
non-trivial check on these models.  We illustrate this below.

We begin by studying certain baryons in the model of Sec.~6.3.  We can
form baryons by introducing dressed quarks \kutsch
\eqn\dress{Q_j=X^jQ;\;\;\;j=0,\cdots k-1.} 
and then contracting the gauge indices on two $SU(N_c)$ epsilon tensors.
\eqn\Sbaryon{B_n^{(n_0,n_1,\cdots,n_{k-1})}=Y^n Q_0^{n_0}Q_0^{n_0}\cdots 
Q_{k-1}^{n_{k-1}}Q_{k-1}^{n_{k-1}}
 ;\;\;n + \sum_{j=0}^{k-1} n_j=N_c}
The total number of baryons is
\eqn\Stotnum{2\sum_{\{n_j\}}{N_f\choose n_0}\cdots{N_f\choose n_{k-1}}=
2{kN_f\choose {N_c - n}}.}
We can similarly form anti-baryons.

Under the duality transformation, baryon operators are mapped to other 
baryon operators in the dual theory.
The mapping is
\eqn\Cbarmap{
B_n^{(n_0,n_1,\cdots,n_{k-1})}
\leftrightarrow
B_{2kN_f + 4 - n}^{({\bar n}_0,{\bar n}_1,\cdots,{\bar n}_{k-1})};\;\;
{\bar n}_j=N_f-n_{k-j-1}}
where in the dual theory the baryons look like 
\eqn\SDbaryon{
B_{2kN_f + 4 -n}^{({\bar n}_0,{\bar n}_1,\cdots,{\bar n}_{k-1})}
=\bar Y^{2kN_f + 4 - n} 
q_0^{{\bar n}_0}q_0^{{\bar n}_0}\cdots q_{k-1}^{{\bar n}_{k-1}} 
q_{k-1}^{{\bar n}_{k-1}}}
The total number of dual baryons is
\eqn\SDtotnumd{2\sum_{\{{\bar n}_j\}}
{N_f\choose {\bar n}_0}\cdots{N_f\choose {\bar n}_{k-1}}=
2{kN_f\choose {\tilde N_c - (2kN_f + 4 -n)}} = 
2{kN_f\choose {kN_f - N_c + n}}.}
which is the same as in the electric theory.
The fact that this mapping is also consistent with all global symmetries 
is another non-trivial test of the proposed duality.

We can consider the flat direction along which a particular baryon,
$\det Y$, gets a vacuum expectation value.  In this direction,
$<Y_{\alpha \beta}> \propto \delta_{\alpha, \beta}$, where $\alpha$
and $\beta$ are gauge indices, breaking the group from $SU(N_c)$ to
$SO(N_c)$. The adjoint field $X$ splits into fields in the symmetric
and the anti-symmetric representations, $X = A + S$.  The field
$\tilde Y$ becomes a field in the symmetric representation of
$SO(N_c)$, $\hat Y$. The superpotential gives a mass to the symmetric
fields $\hat Y$ and $S$, while the field $A$ remains massless, with
superpotential $W=\Tr A^{k+1}$. The low-energy theory is thus an $A_k$
mezzanine model.

Although we have not worked out the 
details of the flat direction corresponding to the dual operator,
the baryon mapping \Cbarmap\ requires that a similar symmetry breaking 
happen in the dual of the high energy theory, through an expectation 
value for the operator $\bar Y^{2kN_f + 4 - N_c} 
q_0^{N_f}q_0^{N_f}\cdots q_{k-1}^{N_f} q_{k-1}^{N_f}$. This
suggests that the dual gauge group breaks from
$SU(3kN_f + 4 - N_c)$ to $SO(2kN_f + 4 - N_c)$ (possibly with other
unbroken confining factors) which would correspond to the dual expected
for an $A_k$ mezzanine model \rlmssosp. 

We turn next to baryons in the model of Sec.~6.4.  As in the case
discussed above, we can introduce dressed quarks and form baryons,
which in this case require only one epsilon tensor since $Y$ is in the
anti-symmetric representation.
\eqn\Abaryon{B_n^{(n_0,n_1,\cdots,n_{k-1})}=Y^n Q_0^{n_0}\cdots 
Q_{k-1}^{n_{k-1}}
 ;\;\;2n + \sum_{j=0}^{k-1} n_j=N_c}

The mapping of baryon operators is
\eqn\Abarmap{
%\eqalign{
B_n^{(n_0,n_1,\cdots,n_{k-1})}
\leftrightarrow
B_{kN_f - 2 - n}^{({\bar n}_0,{\bar n}_1,\cdots,{\bar n}_{k-1})};\;\;
{\bar n}_j=N_f-n_{k-j-1}
%;\cr
%j=0,1,2,\cdots, k-1}
}
where in the dual theory the baryons look like 
\eqn\ADbaryon{
B_{kN_f - 2 -n}^{({\bar n}_0,{\bar n}_1,\cdots,{\bar n}_{k-1})}
=\bar Y^{kN_f - 2 - n} q_0^{{\bar n}_0}\cdots q_{k-1}^{{\bar n}_{k-1}}}

We can consider the flat direction along which a particular baryon,
$\pf Y$, gets a vacuum expectation value.  In this direction, the
group is broken from $SU(N_c)$ to $Sp(N_c)$.  This time it is the
symmetric part of $X$ which is massless, with superpotential $W=\Tr
S^{k+1}$. The low-energy theory is thus an $A_k$ mezzanine model.  As
before, although we have not worked out the details of the flat
direction, the baryon mapping \Abarmap\ requires an expectation value
for the operator $\bar Y^{kN_f - 2 - N_c} q_0^{N_f}\cdots
q_{k-1}^{N_f}$.  This suggests that the dual gauge group breaks to
$Sp(kN_f - 2 - N_c)$ as expected for an $A_k$ mezzanine model
\rlmssosp.

We turn next to baryons in the model of Sec.~7.  We can form baryons
around $Y$ as in Eq. \Abaryon.  The mapping of baryon operators is
\eqn\ACbarmap{
B_n^{(n_0,n_1,\cdots,n_{k-1})}
\leftrightarrow
B_{{k\over 2}(m_f + \tilde m_f) - 
2k - n}^{({\bar n}_0,{\bar n}_1,\cdots,{\bar n}_{k-1})};\;\;
{\bar n}_j=m_f-n_{k-j-1}}
We can also form anti-baryons 
around $\tilde Y$ as in Eq. \Sbaryon.
The mapping of anti-baryon operators is
\eqn\SCbarmap{
\tilde B_n^{(n_0,n_1,\cdots,n_{k-1})}
\leftrightarrow
\tilde B_{k(m_f + \tilde m_f) + 
4k - n}^{({\bar n}_0,{\bar n}_1,\cdots,{\bar n}_{k-1})};\;\;
{\bar n}_j=\tilde m_f-n_{k-j-1}
}

We can consider the flat direction $\vev{\pf Y}$ which breaks the
gauge group to $Sp(N_c)$.  As before $X$ splits into symmetric and
anti-symmetric parts $X = S+A$, with $S$ becoming massive along with
$\tilde Y$, leaving the field $A$ with superpotential $A^{k+1}$.  This
is an $A_k$ orchestra model.  In the dual, the baryon $\bar Y^{{k\over
2}(m_f + \tilde m_f) - 2k - N_c} q_0^{N_f}\cdots q_{k-1}^{N_f}$ gets
an expectation value, suggesting that the low-energy gauge group is
$Sp({k\over 2}(m_f + \tilde m_f) - 2k - N_c)$, which would agree with
the $A_k$ orchestra duality transformation.

Similarly, the flat direction $\vev{\det\tilde Y}$, which breaks the
gauge group to $SO(N_c)$, leaves the field $S$ with superpotential
$S^{k+1}$.  In the dual, the expectation value for the anti-baryon
$\tilde{\bar Y}^{k(m_f +
\tilde m_f) + 4k - N_c} q_0^{N_f}q_0^{N_f}\cdots q_{k-1}^{N_f}
q_{k-1}^{N_f}$ suggests the low-energy gauge group is $SO(k(m_f +
\tilde m_f) + 4k - N_c)$, which would again agree with the $A_k$
orchestra duality transformation.

\newsec{Summary and Conclusions.}

We have presented a number of new examples of duality, which are
generalizations of the $D_{k+2}$ models of Ref.~\brodie.  The pattern
of new examples resembles the pattern of the $A_k$ models
\refs{\kutsch,\kenisosp,\rlmssosp,\ilslist} as is clear from comparing
\figAk\ and \figDk.  
The dualities of the $D_{k+2}$ and $A_k$ models have much in common,
such as the fact that they
relate pairs of theories of similar type, the presence of gauge
singlet mesons in the magnetic superpotentials, and the existence at
all self-dual points of marginal operators which take the form of
meson mass terms.

However, despite the similarities of these patterns, there
are many differences as well, and a number of puzzles remain.

First, although the traditional ADE classification of groups
and singularities contains $D_{k+2}$ for all positive $k$,
the present examples of duality seem only to allow $k$ odd, except for the 
chiral balcony theory where the algebra is different.  This is 
more than just a mathematical oddity; it is a real problem
for the duality.   Were our understanding of these
dualities complete, we would know the mapping of the
operator $X^{2p+1}$ to the dual theory.  We could then perturb
the electric superpotential by $X^{2p+1}$ and the magnetic superpotential
by the image of this operator, and the electric theory would then
flow to a $D_{2p+2}$ model with the magnetic theory flowing to its
dual.  We would thereby derive the duality for $k$ even.  But this
approach fails, showing that we do not understand the mapping
for the operators $X^{2p+1}$. It is especially confusing that
this issue does not arise for the balcony $D_{k+2}$ model.  Why
this theory is fundamentally different is not yet clear to us.

Second, as we have seen, the field $Y$ may be chosen to be
any one of the standard two-index tensors of $SU$, $SO$, or
$Sp$, but the field $X$ must be in the adjoint of $SU$, the
symmetric tensor of $SO$, or the anti-symmetric tensor of $Sp$.
These three fields appear in the orchestra models of the $A_k$
series \refs{\kutsch,\kenisosp,\ilslist}.  For other choices
of $X$ the superpotential term $XY^2$ cannot be written, and
no obvious generalization seems to work.  Note that in the
$A_k$ series many generalizations of the original $X^{k+1}$ 
superpotential appear in the list of dual models.
Why $X$ is so restricted in the $D_{k+2}$ models is unknown.

Third, unlike the $A_k$ series, there do not seem to be any loge
models of the form $G\times G'$ which confine to form the models of
the $D_{k+2}$ series.  We do on the other hand find models of the form
of a product of $SO$ or $Sp$ groups which generalize the stage 
model with $SU\times SU$ gauge group found in Ref.~\brodie.  All
the $D_{k+2}$ models with gauge group $G$ flow under certain
superpotential perturbations to the stage models with gauge
group $G\times G$, except for the balcony model, which flows to
copies of itself (with smaller $k$) under such perturbations.

Fourth, we have not yet found any sign of models
corresponding to the singularities for $E_6$, $E_7$ or $E_8$.
Notice however that $E_4=A_4$ and $E_5=D_5$ do appear in our list.
The significance of this is again unclear.

Fifth, the patterns of the dual groups as listed on the right-hand
side of \figAk\ do not closely resemble those of \figDk , despite 
the fact that the $A_3$ and $D_3$ cases agree.  There
are clear patterns in both cases; but why should they be so different
from one another?  Are there other classes of models which, once 
discovered, will show that the organizing plan that we have used 
is misleading?

The more basic issues raised by this work point toward a number
of areas for additional research.  What role is this matrix generalization
of singularity theory playing in duality?  How is it connected with
the new concepts of matrix geometry which arise in the context
of D-branes \M\ ?  Are there D-brane constructions of these theories
in which the duality transformations might be manifest, or if not
manifest at least related to known string-associated dualities? Are
marginal operators important in understanding the nature of
or source of duality, or are they just an interesting sidelight?
We expect there will be many interesting discoveries to come.

\bigskip

{\bf Acknowledgements}

\medskip

We thank Ken Intriligator for discussions.  M.J.S. thanks the Aspen
Center for Physics where this work was begun.  J.H.B. was supported by
National Science Foundation grant PHY90-21984.  M.J.S. was supported
by National Science Foundation grant NSF PHY-9513835 and by the WM
Keck Foundation.

\listrefs
%\listfigs
\end